\documentclass[preprint2,times,tighten]{aastex6}

\usepackage{amsmath,amstext}
\usepackage[figure,figure*]{hypcap}
\usepackage{newtxmath}

\usepackage{graphicx}
\usepackage{amssymb}

\shorttitle{GASP II. Kinematics of the jellyfish galaxy JO201}
\shortauthors{Bellhouse et al.}

\begin{document}
\title{GASP II. A MUSE view of extreme ram-pressure stripping along the line of sight: \\
kinematics of the jellyfish galaxy JO201}

\author{C. Bellhouse\altaffilmark{1,2}}
\author{Y.~L. Jaff\'e \altaffilmark{2}}
\author{G. K. T. Hau \altaffilmark{2}}
\author{S.~L. McGee\altaffilmark{1}}
\author{B.~M. Poggianti\altaffilmark{3}}
\author{A. Moretti\altaffilmark{3}}
\author{M. Gullieuszik\altaffilmark{3}}
\author{D. Bettoni\altaffilmark{3}}
\author{G. Fasano\altaffilmark{3}}
\author{M. D'Onofrio\altaffilmark{3,7}}
\author{J. Fritz\altaffilmark{4}}
\author{A. Omizzolo\altaffilmark{3,6}}
\author{Y.-K. Sheen\altaffilmark{8}}
\author{B. Vulcani\altaffilmark{5}}

\email{cbellhou@eso.org}
\thanks{Present address: European Southern Observatory, Alonso de Cordova 3107, Vitacura, Santiago, Chile}
\altaffiltext{1}{University of Birmingham School of Physics and Astronomy, Edgbaston, Birmingham, England}
\altaffiltext{2}{European Southern Observatory, Alonso de Cordova 3107, Vitacura, Casilla 19001, Santiago de Chile, Chile}
\altaffiltext{3}{INAF - Astronomical Observatory of Padova, vicolo dell'Osservatorio 5, I-35122 Padova, Italy}
\altaffiltext{4}{Instituto de Radioastronom\'ia y Astrof\'isica, Morelia, Mexico}
\altaffiltext{5}{School of Physics, The University of Melbourne, VIC 3010, Australia}
\altaffiltext{6}{Specola Vaticana, 00120, Vatican City State}
\altaffiltext{7}{Dipartimento di Fisica \& Astronomia “Galileo Galilei”, Universit\`a di Padova, vicolo dell’ Osservatorio 3, IT 35122, Padova, Italy}
\altaffiltext{8}{Korea Astronomy and Space Science Institute, 766, Daedeokdae-ro, Yuseon-gu, Daejon, 34055, Korea}

\date{Last updated \today; in original form.}

\begin{abstract}

This paper presents a spatially-resolved kinematic study of the jellyfish galaxy JO201, one of the most spectacular cases of ram-pressure stripping (RPS) in the GASP (GAs Stripping Phenomena in Galaxies with MUSE) survey. By studying the environment of JO201, we find that it is moving through the dense intra-cluster medium of Abell 85 at supersonic speeds along our line of sight, and that it is likely accompanied by a small group of galaxies. Given the density of the intra-cluster medium and the galaxy's mass, projected position and velocity within the cluster, we estimate that JO201 must so far have lost $\sim 50$\% of its gas during infall via RPS.  
The MUSE data indeed reveal a smooth  stellar disk, accompanied by large projected tails of ionised ($\mathrm{H}{\alpha}$) gas, composed of 
kinematically cold (velocity dispersion $<40 $km~s$^{-1}$) star-forming knots and very warm ($>100 $km~s$^{-1}$) diffuse emission which extend out to at least ${\sim} 50$~kpc from the galaxy centre. 
The ionised $\mathrm{H}{\alpha}$-emitting gas in the disk rotates with the stars out to $\sim 6$~kpc but in the disk outskirts becomes increasingly redshifted with respect to the (undisturbed) stellar disk. The observed disturbances are consistent with the presence of gas trailing behind the stellar component, resulting from intense face-on RPS happening along the line of sight. Our kinematic analysis is consistent with the estimated fraction of lost gas, and reveals that stripping of the disk happens outside-in,  causing shock heating and gas compression in the stripped tails.

\end{abstract}


\section{Introduction}

The evolution of galaxies is driven by a combination of internal and environmental processes. 
Important discoveries from the past decades supporting this idea include the observation that galaxy morphology and colour correlate with both environmental density and galaxy mass \citep[e.g.][]{1980ApJ...236..351D,2006MNRAS.373..469B, 2010ApJ...721..193P}.
Internal processes affecting the evolution of galaxies include stellar feedback from high-wind stars such as Wolf-Rayet and OB type stars, supernova feedback, and the effects of AGN activity \citep{2005ARA&A..43..769V,2014MNRAS.444.3894H}. Environmental processes can be separated into two categories: gravitational and hydrodynamical.
Among the gravitational effects are tidal interaction between galaxies, common in moderately crowded environments \citep{1951ApJ...113..413S, 1983ApJ...264...24M,Toomre1977,Tinsley1979,Mihos1994,Springel2000}.
In higher-mass clusters, the galaxies may also be subjected to perturbation by the gravitational potential of the cluster as a whole \citep{Byrd1990,Valluri1993}, giving rise to inflows as well as boosting star formation within the nuclear and disk regions. Over extended periods of time, cluster galaxies can also be affected by cumulative high-speed close-approach encounters with other cluster members   \citep[\textit{harassment;}][]{1996Natur.379..613M, 1998ApJ...495..139M}. 
Hydrodynamical interactions between galaxies and  the intracluster medium (ICM) can remove their halo gas via \textit{starvation/strangulation} \citep{Larson1980,Balogh2000}, or their interstellar gas via ram-pressure stripping \citep[RPS; ][]{1972ApJ...176....1G, Faltenbacher2006, Takeda1984}. Additional mechanisms such as thermal evaporation have also been proposed \citep{Cowie1977}. 
We refer to \citet{2006PASP..118..517B} for a review of all the environmental effects.  In this paper, we focus on the effect of RPS on the gas loss of cluster galaxies.

In a seminal paper, \citet[][]{1972ApJ...176....1G} showed that the ram-pressure exerted by the ICM on an infalling galaxy is proportional to the medium density  and the square of the velocity of the galaxy relative to it $(\mathrm{P}_{\rm ram} \propto \rho v^2)$. 
The galaxy will lose its gas if ram-pressure overcomes the limiting force per unit area,  which depends on the surface density.
This description, although simplified,  reproduces very well the observed  gas deficiencies of cluster galaxies \citep[see e.g.][]{Chung2007,2015MNRAS.448.1715J}. Recent hydrodynamical simulations have shown however that the efficiency of RPS is a function of the inclination, with face-on encounters being typically more efficient at stripping than edge-on or inclined encounters \citep{1999MNRAS.308..947A,2000Sci...288.1617Q,2001ApJ...561..708V}.
Some studies further suggest that gas removal by RPS may be accompanied by an enhancement in
the star formation activity in the galaxy, as thermal instabilities and turbulent motions within the galaxy provoke collapse within cold clouds  \citep{1991MNRAS.248P...8E, 2003ApJ...596L..13B}. 
In fact, during the stripping phase, galaxies can 
exhibit largely disturbed optical or ultraviolet tails composed of young stars formed in-situ in the stripped gas \citep[][]{Kenney1999,Cortese2007,Yoshida2008,Hester2010,Yagi2010,Smith2010,Owers2012,Kenney2014,Ebeling2014,Fumagalli2014,Rawle2014,McPartland2016}, for  which they are often referred to as ``jellyfish" galaxies.

%
After the active stripping period has passed, and a significant portion of the galaxy's gas has been lost to the ICM, the quenching of its star formation becomes inevitable
\citep{1998ApJ...509..587F, 1999ApJ...516..619F, Haines2015,2016MNRAS.tmp..769J,Vollmer2012}.

The most convincing examples of RPS at play come from observations of  neutral atomic Hydrogen gas (HI) in cluster galaxies, that display ``tails'' of gas trailing from the stellar continuum and reduced gas fractions that have convincingly demonstrated the efficiency of RPS \citep[e.g.][Yoon et~al.~accepted]{Haynes1984,VerdesMontenegro2001,Cayatte1990,2004AJ....127.3361K, 2009AJ....138.1741C,Cortese2010}. 
Molecular gas studies show that while this component is vulnerable to RPS, it is much less affected than atomic gas \citep{Kenney1989,Boselli1997,Boselli2014}.
An increasing number of studies of the ionised gas in jellyfish galaxies is also helping to  understand the impact of RPS on star formation  \citep{Gavazzi2002b,Yagi2010,Fossati2012,Boselli2016}, with integral-field observations being particularly insightful \citep{Merluzzi2013,Fumagalli2014,2016MNRAS.455.2028F}.
Finally, jellyfish galaxies have been successfully reproduced by hydrodynamical simulations of RPS
\citep[][]{Roediger2007,Tonnesen2009,Tonnesen2010,Tonnesen2012}.

We present results from a new 
MUSE \citep[Multi Unit Spectroscopic Explorer;][]{2010SPIE.7735E..08B} survey at the VLT called GAs Stripping Phenomena in galaxies with MUSE (GASP\footnote{\url{web.oapd.inaf.it/gasp}}), that aims to provide a detailed systematic study of gas removal processes in galaxies (see Poggianti et al. submitted; Paper I).
GASP is observing over 100 galaxies from the sample of jellyfish candidates of \citet{2016AJ....151...78P}, who visually selected galaxies with signatures of stripping from optical images of the WIde-field Nearby Galaxy-cluster Survey (WINGS) \citep{Fasano2006, Moretti2014}, its extension OmegaWINGS \citep[][]{Gullieuszik2015,Moretti2017}, and the Padova-Millennium Galaxy and Group Catalogue \citep{Calvi2011}. The sample thus span a wide range of stellar masses $(10^{9.2}-10^{11.5} \mathrm{M}_\odot)$ and environment (host dark matter haloes of $10^{11}-10^{15} \mathrm{M}_\odot$).

In this paper we present a kinematical analysis of one of the most spectacular jellyfish galaxies in GASP, JO201 (WINGSJ004130.30-091546.1, also known as PGC 2456 and KAZ 364, RA 00:41:30.325, Dec -09:15:45.96), a heavily ram-pressure stripped galaxy in the massive galaxy cluster Abell 85. In section \ref{sec:env} we summarise the properties of JO201 and present a study of its local and global environment. 
In section \ref{sec:obs} we describe the GASP MUSE observations, the data reduction process and the corrections used for Galactic extinction. Section \ref{sec:linefits} focuses on the H$\alpha$ emission-line fitting processes and shows the resulting H$\alpha$ emission maps. 
Section \ref{sec:kinematics} presents the stellar and ionised gas kinematics of the galaxy, including velocity and velocity dispersion maps, as well as rotation curves. Finally, in section \ref{section:conclusion} we summarise all of the results and draw conclusions.

Unless otherwise stated, throughout the paper we adopt a Chabrier initial mass function \citep[IMF;][]{Chabrier2003}, and a concordance $\Lambda$CDM cosmology of $\Omega_\mathrm{M}=0.3$, $\Omega_\Lambda=0.7$, $\mathrm{H}_0=70\mathrm{km}\,\mathrm{s}^{-1}\mathrm{Mpc}^{-1}$.

\section{JO201 and its environment} \label{sec:env}

\begin{figure}\centering
\includegraphics[width=0.5\textwidth]{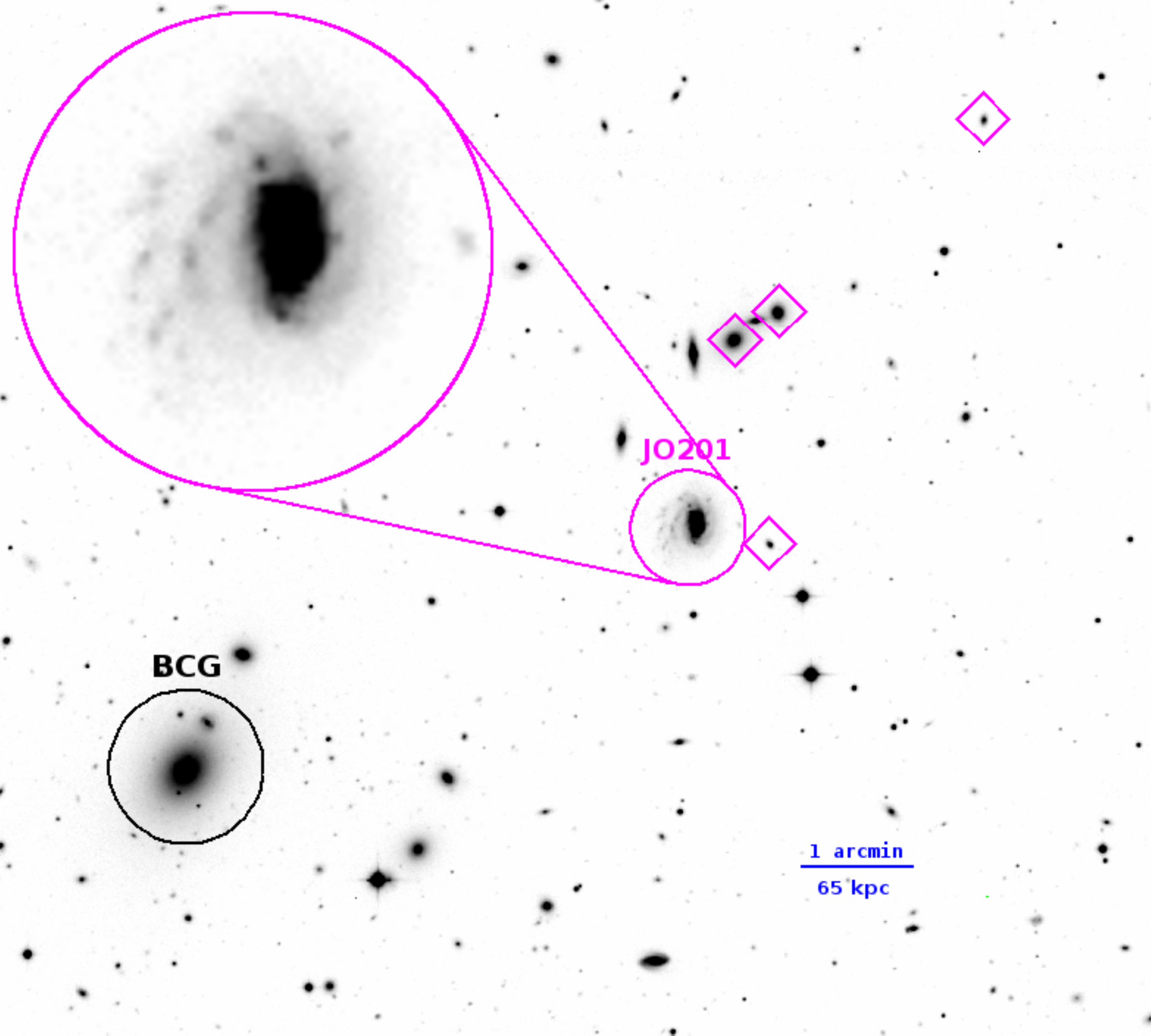}
\caption{WINGS V-band image showing the central part of Abell 85. Highlighted are the BCG (black circle), the jellyfish galaxy JO201 (magenta circle and inset) and its possible group companions (magenta diamonds), selected from their kinematical deviations as described in the text. The distance between JO201 and the BCG is $\sim$5.6 arcmin (360~kpc at the redshift of the cluster). East is to the left and north to the top.\label{figure:A85map}}
\end{figure}

\begin{figure*}
\includegraphics[width=0.5\textwidth]{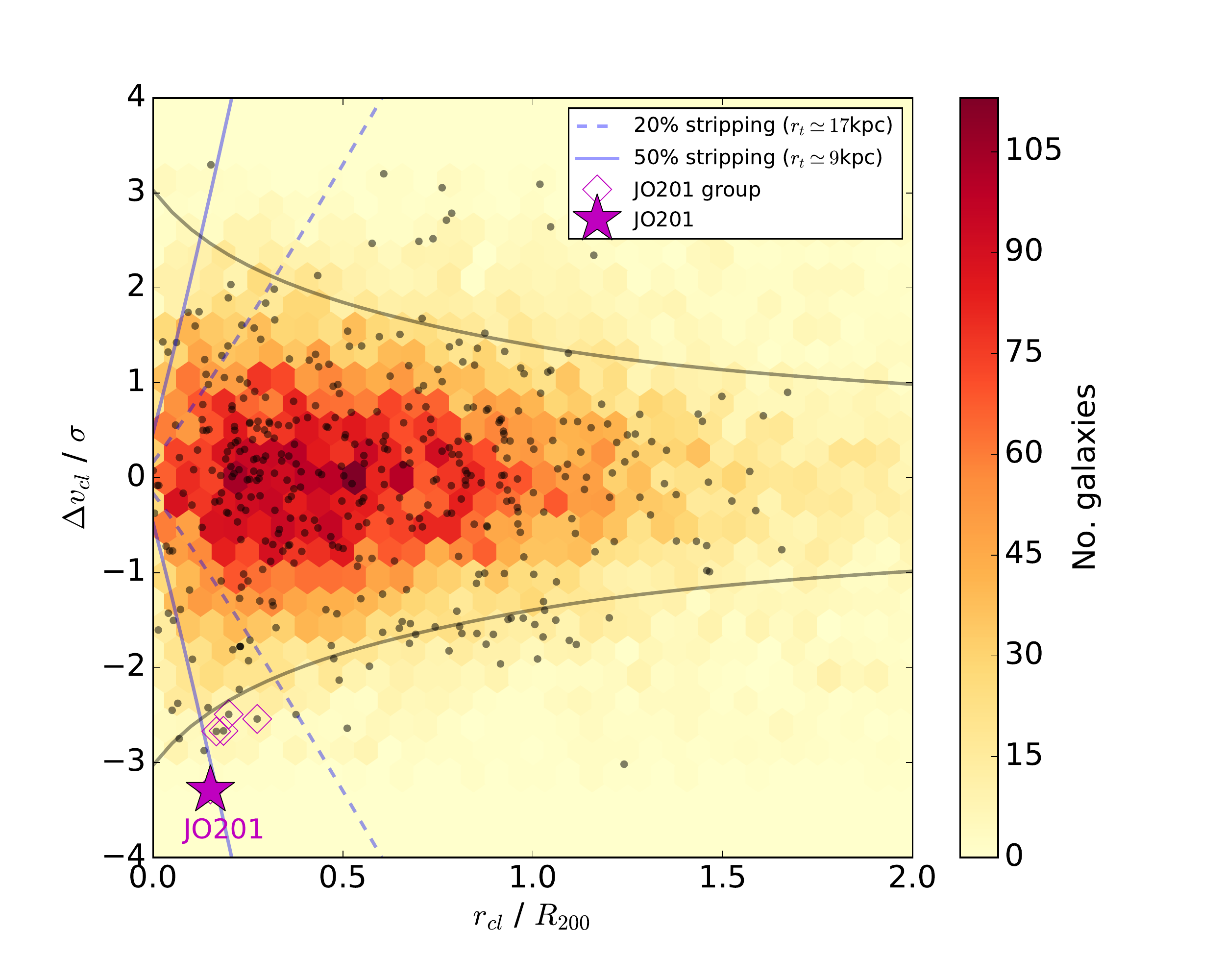}
\includegraphics[width=0.5\textwidth]{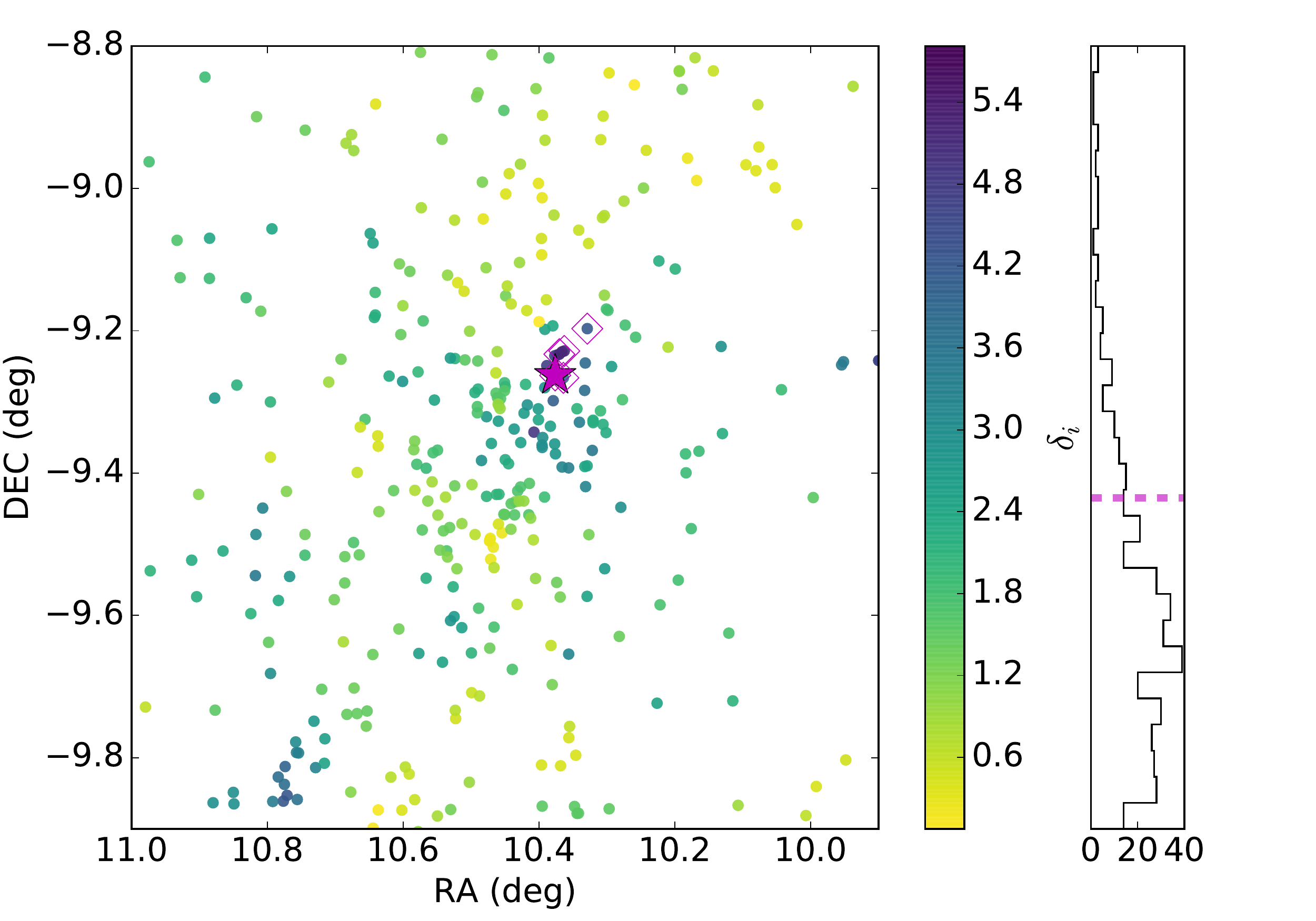}
\caption{Left: Projected phase-space diagram of all the OmegaWINGS clusters (background density plot), where galaxies in Abell 85 are highlighted with gray circles. The gray curves represent the escape velocity in a \citet*{1996ApJ...462..563N} halo profile. The dashed and solid blue curves represent RPS models of JO201 by A85's ICM, in which 20\% or 50\% of the total gas mass in the disk has been removed (see text for details). Right: Dressler-Shectman ``bubble'' plot, showing the spatial distribution of galaxies in Abell 85, colour-coded by their velocity deviations $\delta_{i}$. The magenta dashed line on the histogram shows $\delta_i = 2.5$, above which a galaxy may be considered to be part of a group or substructure.
In both panels JO201 is marked with a magenta star and its possible group companions with open magenta diamonds.}\label{figure:Env2}
\end{figure*}

JO201 is perhaps the most convincing case of gas stripping in the sample of 419 (344 cluster and 75 field) jellyfish candidates of  \citet{2016AJ....151...78P}. Its morphology is that of a spiral galaxy with tails of material to one side and its total stellar mass is 
$3.55_{-0.32}^{+1.24} \times 10^{10} \mathrm{M}_{\odot}$. 

X-ray studies have classified  JO201 as a Seyfert galaxy \citep{Durret2005}. Our MUSE observations confirm the presence of an AGN (see e.g. broadened emission-lines at the centre of the galaxy in Section~\ref{sec:linefits}), but we leave the study of the AGN activity in JO201 for a following paper (Bellhouse et al. in preparation). 

JO201 resides in Abell 85, a massive cluster with a velocity dispersion of $\sigma_{cl}=982
\pm 55$km~s$^{-1}$  ($M_{200}=1.58 \times 10^{15} \mathrm{M}_{\odot}$), 
at a redshift of $z_{\rm cl}=0.05586$ \citep{Moretti2017}.  This cluster is part of OmegaWINGS and therefore has wide-field OmegaCam imaging (${\sim 1} \mathrm{deg}^2$) and spectroscopy from the Anglo-Australian Telescope.
JO201 is located very close to the brightest cluster galaxy (BCG), with a 
projected radial distance ($r_{cl}$) of only 360~kpc ($0.15 \times R_{200}$), as can be seen in Figure~\ref{figure:A85map}. 
Moreover, JO201 has a very high line-of-sight  velocity with respect to the mean velocity of the cluster 
($|\Delta v_{cl}|= 3363.7$km~s$^{-1} = 3.4 \times \sigma_{cl}$). In fact, it is outside the velocity cut originally considered for cluster membership 
\citep[$3\times \sigma_{cl}$; Moretti et al. accepted;][]{Cava2009}.

The left panel of Figure~\ref{figure:Env2} shows the projected position vs. velocity phase-space diagram of galaxies in Abell 85. All the cluster members with a spectroscopic redshift in Abell 85 are over-plotted (gray filled circles) on top of the distribution of all OmegaWINGS clusters stacked together (hex plot), and lie mostly within the trumpet-shaped region defined by the escape velocity of a \citet*{1996ApJ...462..563N} halo profile (grey curves). In this diagram, JO201 (magenta star) lies very close to the central region of the cluster in the ``envelope" of the trumpet, which suggests that this galaxy is likely falling into the cluster for the first time. Its high peculiar velocity combined with its small $r_{cl}$ also indicates a highly radial orbit. It is worth noting that JO201 is the most central and fast-travelling cluster jellyfish galaxy from the sample of \citep[][see Jaff\'e et al. in preparation]{2016AJ....151...78P}.    

Its location in phase-space suggests that the galaxy is falling into the cluster from behind and moving towards the observer.  Although the line-of-sight component probably dominates the velocity vector, the galaxy's  projected tails pointing eastwards (close to the direction of the BCG; see Figure~\ref{figure:A85map}), further suggest that the velocity vector must have some inclination with respect to the line-of-sight.

We quantify the ram-pressure intensity exerted on JO201 by the ICM in the same manner as in \citet{2015MNRAS.448.1715J}. We utilize the cluster gas density distribution of Abell 85 presented in \citet{2009A&A...495..379B}, computed from a hydrostatic-isothermal $\beta$-model \citep{Cavaliere1976} as: 
\begin{equation}
\rho(r) = \rho_0[1+(r_{cl}/r_c)^2]^{-3\beta/2},
\end{equation}
where the core radius is $r_{c}=82$ kpc, the central density $\rho_0=0.0257$cm$^{-3}$, and $\beta=0.532$ 
\citep[values from][]{Chen2007}. The ram-pressure exerted on JO201 can be computed following 
\citet{1972ApJ...176....1G}'s equation: $P_{\rm ram} = \rho_{\rm ICM} \times v^{2}$, 
where  $\rho_{\rm ICM}$ is the density of the ICM at the galaxy's position in the cluster and $v$ the galaxy's velocity relative to the ICM. We adopt the projected cluster-centric distance ($r_{cl}$) and the line-of-sight differential velocity of the galaxy ($|\Delta v_{cl}|$) and obtain $P_{\rm ram}(\mathrm{JO201}) = 1.24 \times 10^{-11} N m^{-2}$. Note that due to the mass of A85 and the extreme speed of JO201, this value is over an order of magnitude higher than the estimated pressures computed for Virgo galaxies with HI tails in \citep{Chung2007}. 

We can then compare ram-pressure exerted by the ICM on the infalling galaxy with the anchoring self-gravity provided by the galaxy, defined as: 
\begin{equation}
\Pi_{\rm gal}=2 \pi G \Sigma_{\rm s}\Sigma_{\rm g}, 
\label{eq:pi}
\end{equation}
where $\Sigma_{\rm s}$ and $\Sigma_{\rm g}$ are the density profiles of the stellar and gaseous exponential disks respectively, that can be expressed as: 
\begin{equation}
\Sigma=  \left(\frac{M_{d}}{2 \pi r_d^2}\right)  e^{-r/r_d},
\label{eq:sigma0}
\end{equation}

where $M_{d}$ is the disk mass, $r_d$ the disk scale-length and $r$ the radial distance from the center of the galaxy. 
For the stellar component of JO201 we adopted a disk mass  $M_{d,stars}=3.55\times10^{10}M_{\odot}$,
and a disk scale-length $r_{d,stars} = 5.56 $~kpc, obtained by fitting the light profile of the galaxy from the V-band WINGS image\footnote{We note that, although the light profile fitting reveals that JO201 is dominated by a disk component, there is a non neglible bulge component (20\% of the light) that we ignore when estimating the ram-pressure stripping intensity across the galaxy. The bulge component however is confined to the central part of the galaxy. We also note that the value $R_d$ obtained from V-band photometry is consistent with fits made to the stellar continuum (near H$\alpha$) using the MUSE datacube.}.   
For the gas component we assumed a total mass $M_{d,gas} = 0.1 \times M_{d,stars}$,
and scale-length $r_{d,gas} = 1.7 \times r_{d,stars}$  \citep{2006PASP..118..517B}.

These values yield 
$P_\mathrm{\rm ram} =   0.008 \times \Pi_{\rm gal}$ 
%
at the centre of the galaxy, which is not enough to strip the inner gas. 
The condition for stripping ($P_{\rm ram}$/$\Pi_{\rm gal} > 1$) is met only when considering the galaxy is partially stripped (outside-in) down to a given ``truncation radius", $r_t$.

By combining the stripping condition with equations~\ref{eq:pi} and~\ref{eq:sigma0} at $r=r_t>0$, we find find that at the location of JO201 in phase-space, $P_{ram}$ should have stripped the gas in JO201
%
down to $r_t\simeq$9~kpc, which is equivalent to $\sim$50\% of the total gas mass stripped, estimated from the equation for remaining gas mass:  
\begin{equation}
f=1+\left[e^{-r_t/r_d}\left(\frac{-r_t}{r_d} -1\right)\right] \label{eq:frac}
\end{equation}
This is shown in the phase-space diagram of Figure~\ref{figure:Env2} (left), where   JO201 (magenta star) sits on top of the line corresponding to 50\% gas stripping of JO201 in A85 (solid blue lines). We also plot the region corresponding to 20\% of gas stripping (dashed blue lines; $r_t\simeq 17$~kpc) as a reference to the trajectory and stripping history of JO201 in phase-space. As we will see in Section~\ref{sec:RC}, the extent of the non-stripped $\mathrm{H}\alpha$ emitting disk ($r\gtrsim 6$~kpc along the major axis) is comparable to the estimated $r_{t}$. Note however that $r_{t}$ is particularly difficult to measure in this galaxy due to the direction of stripping (see Sections~\ref{sec:linefits} and~\ref{sec:kinematics}). 

We further study the local environment of JO201 by investigating the presence of substructures within Abell 85. We perform a Dressler-Shectman test  \citep{1988AJ.....95..985D}, that consists of computing  individual galaxy deviations by comparing the \textit{local} (nearest neighbours) velocity and velocity dispersion for each galaxy with the \textit{global} (cluster) values. 
We characterise the cluster by its  mean velocity ($\bar{v}_{\rm cl}$), velocity dispersion  ($\sigma_{\rm cl}$), and the total number of cluster members ($N_{\rm mem}$). Then, for each galaxy with $|\Delta v_{cl}| < 4\times \sigma_{cl}$ (this cut includes galaxies with borderline membership, like JO201),  we select a subsample of galaxies containing the galaxy $i$, plus its nearest 10 neighbors, and compute their mean velocity $\bar{v}^{i}_{\rm local}$ and velocity dispersion $\sigma^{i}_{\rm local}$. From these, we compute the individual galaxy deviations  $\delta_{i}$, following:

\begin{equation}
\label{deltai}
\delta^{2}_{i}=\left(\frac{10+1}{\sigma^{2}_{\rm cl}}\right) \left[ (\bar{v}^{i}_{\rm local}  -  \bar{v}_{\rm cl})^{2} + (\sigma^{i}_{\rm local}  - \sigma_{\rm cl})^{2}  \right]
\end{equation}

One way to investigate whether the cluster is substructured is through the ``critical value" method, that compares $\Delta = \sum(\delta_i)$ with $N_{\rm mem}$. For A85, $\Delta / N_{\rm mem} = 1.7$, and, as explained in \citet{1988AJ.....95..985D}, a value $>1$ is found in clusters with significant amounts of substructure. 

The computed individual $\delta_{i}$ values are shown in the right panel of Figure~\ref{figure:Env2}, where groups of galaxies with high velocity deviations ($\delta_i > 2.5$, as indicated by the dashed line in the right-most histogram) are considered potential groups. According to this criteria several substructures are visible (e.g. at the centre and south-east).
Our results are consistent with previous optical studies of A85 such as \citet{BravoAlfaro2008}. Moreover, the high level of substructure found in A85 is in agreement with the findings of \citet{Ichinohe2015}, who studied the cluster's X-ray emission and concluded that the cluster has undergone a merger event in the past, in addition to at least two currently ongoing mergers with galaxy groups.

To test whether JO201 is part of a substructure, we select galaxies in the vicinity of JO201 with $\delta_i > 2.5$ and study their velocity distribution. We find that 4 of these galaxies are likely forming a group with JO201, since they have similar line-of-sight velocities (see left panel of Figure~\ref{figure:Env2}). These galaxies are highlighted in Figures \ref{figure:A85map} and \ref{figure:Env2} with open blue and magenta diamonds respectively.

In summary, JO201 is a massive spiral falling into the massive galaxy cluster Abell 85 within a small group of galaxies.  Its projected position and velocity within the cluster suggests that  it is crossing the cluster at supersonic speed in a radial orbit. The galaxy's motion through the cluster invokes ram-pressure on its disk, which has caused significant gas stripping ($\sim$50\% of total gas mass). In the following sections we study the effects of RPS on this galaxy by comparing the distribution and kinematics of ionised H$\alpha$ gas with that of stars.

\section{Data}\label{sec:obs}

\begin{figure}\centering
\includegraphics[width=0.45\textwidth]{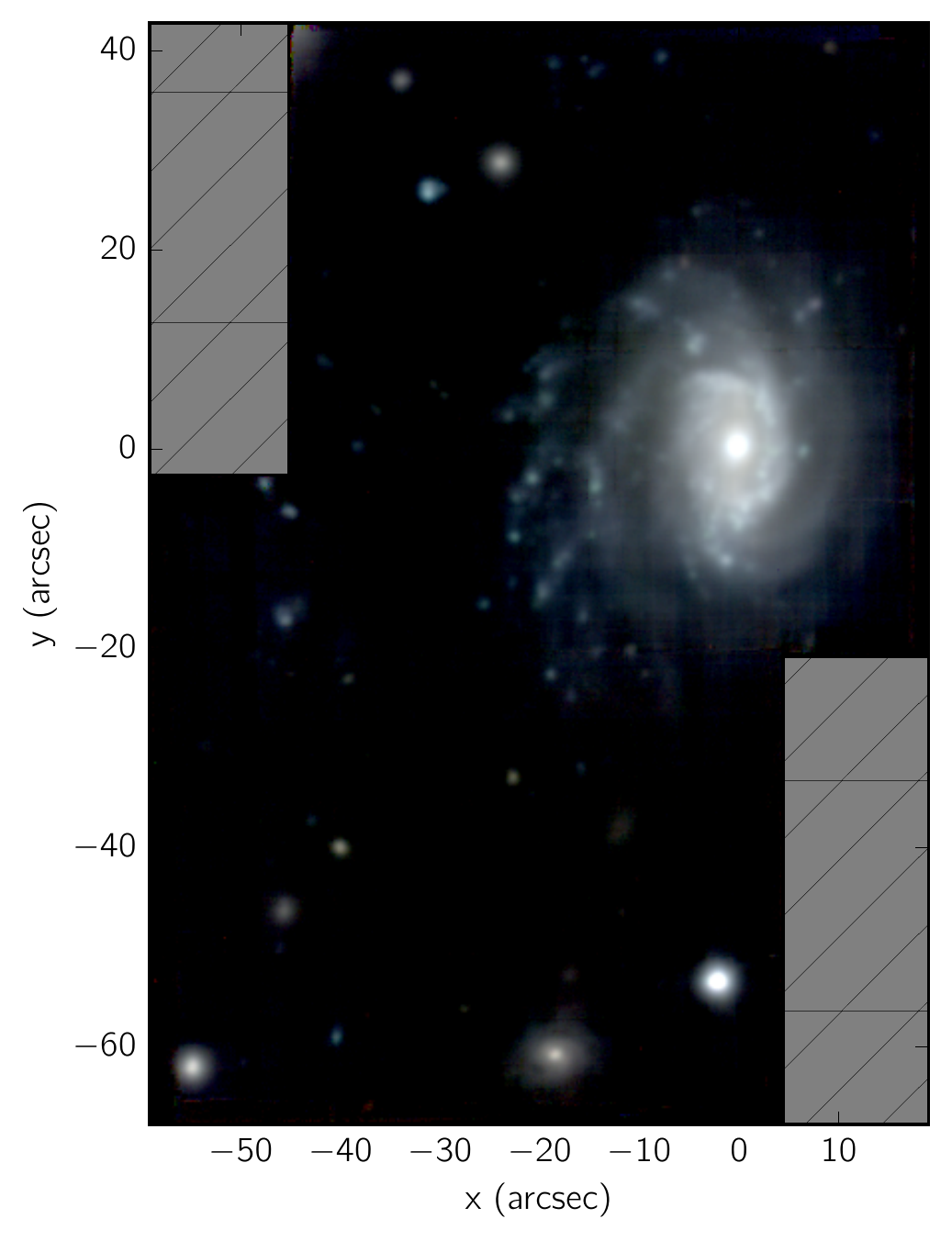}
\caption{RGB image from JO201 GASP observation produced using $1000\AA$ slices integrated from the datacube (B:5000-6000\AA, G:6000-7000\AA, R=7000-8000\AA). East is to the left and north to the top. A series of bright knots to the east of the galaxy are seen connected by ``bridges" of diffuse emission. The asymmetry of the disk and brightening of the western edge is suggestive of a bow-shock increasing the star-formation activity in the leading edge. On the far left are bright knots which are far removed from the galaxy and likely to have been stripped at a much earlier stage.}\label{figure:rgbJO201}
\end{figure}

\subsection{MUSE Observations}

JO201 was observed on 2015-12-17, with photometric conditions and image quality of ${\sim}0.7$" FWHM, as measured from the stars in the Slow-Guiding System surrounding the MUSE field of view (see MUSE user manual\footnote{\url{http://www.eso.org/sci/facilities/paranal/instruments/muse/doc.html}}). A total of eight 675 second exposures spanning 2 adjacent fields were observed with the Nominal Mode, with an instrument rotation of 90 degrees and a small spatial offset between exposures as recommended by the MUSE User Manual. An internal illumination flat field was taken at the beginning of the observations for the purpose of illumination correction. A spectrophotometric standard star GD-71 was also observed right after the science target, for flux-calibration and telluric-correction. Standard calibration files as per the MUSE Calibration Plan were also taken. 

\begin{figure}\centering
\includegraphics[width=0.48\textwidth]{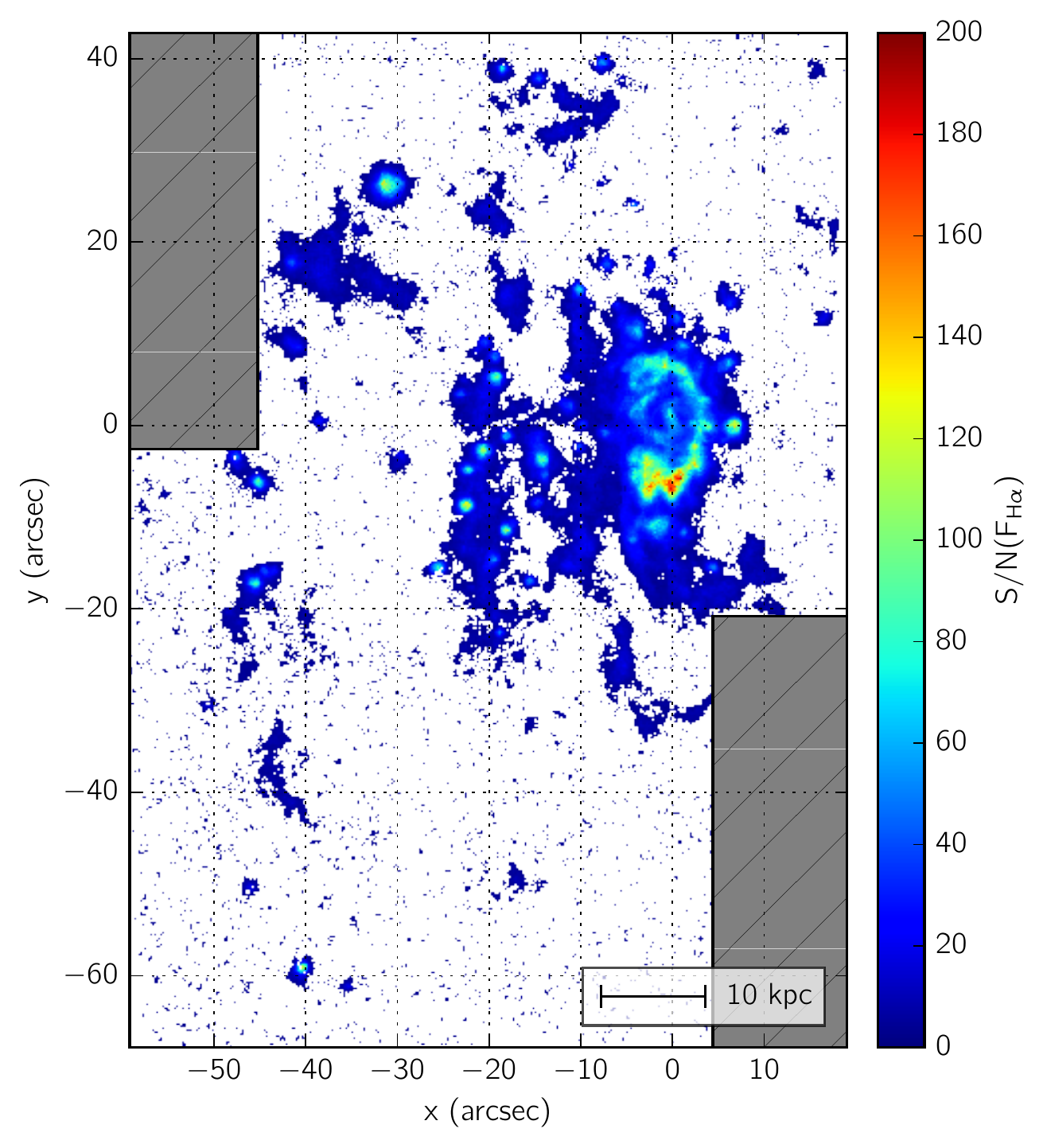}
\caption{$\mathrm{H}\alpha$ signal-noise-ratio map calculated from the line fits to the MUSE data. The higher S/N ratios are seen in the areas of brightest $H\alpha$ emission in the western edge of the galactic disk and in the knots located eastwards of the disk. All fitted spaxels are shown with no signal noise cutoff applied in this figure.
}\label{figure:snJO201}
\end{figure}

\subsection{Data reduction}

For data reduction, we followed the standard reduction procedures as described in the pipeline manual \citep[][\url{www.eso.org/sci/software/pipelines/muse}]{2010SPIE.7735E..08B}, with minor exceptions. We used custom scripts for organising and preparing the raw data, which were then fed to ESOREX recipes v3.12, MUSE pipeline version 1.2.1. As the data have sufficient sky coverage, the sky was modelled directly and subtracted from individual frames using the 20\% pixels with the lowest counts. After wavelength calibration using arc lamp exposures, the final wavelength adjustments were made using sky emission lines. The final, flux-calibrated data cube was generated by lining up and combining the individual frames using sources in the white light images. The final image quality is 0.8" FWHM in the reduced cube.

The full GASP data reduction process is described in greater detail in Paper I.

An RGB image produced using the reduced MUSE cube is shown in Figure \ref{figure:rgbJO201}. Visible in the image are a series of bright knots to the East of the galaxy connected by ``bridges" of diffuse emission as well as slight morphological asymmetry in the disk. The western side of the disk shows a bright ridge of emission along an arc shaped pattern, suggestive of a bow-shock increasing the star-formation activity in the leading edge. On the far left of the image are bright knots which are far from the disk of the galaxy and are likely to have been stripped at a much earlier stage.

\subsection{Galactic Extinction Correction}

The observed spectra are subject to reddening by dust in the Milky Way, which must be accounted for prior to any analysis of the emission line fluxes. For the correction to be made, the amount of dust visible in the observed line of sight must be quantified. We estimated the contribution of dust using values from the galactic dust extinction and reddening tool on the NASA/IRSA infrared science archive, which uses measurements from \citet{Schlafly2011} to quantify the dust distribution and give an extinction estimate across the cube.
The extinction obtained of $\mathrm{E}(\mathrm{B}-\mathrm{V})= 0.0319$, $\mathrm{A}_\mathrm{V}=0.0987$ was used in the Cardelli extinction law \citep{1989ApJ...345..245C} to obtain a correction which was applied to the spectrum of each spaxel across the cube. The corrected cube was then used in all subsequent analyses.

\section{Emission Line Fitting}\label{sec:linefits}

By fitting model profiles to the emission lines, the spatial distribution of different emission lines as well as the peculiar velocity and velocity dispersion of the emitting gas can be derived. In order to fit emission lines, the IDL custom code \textsc{kubeviz} \citep{2016MNRAS.455.2028F}, written by Matteo Fossati and David Wilman, was used. \textsc{kubeviz} fits a set of predefined emission lines across the spectrum using models with up to two gaussian components or using moments.

\subsection{Single Component Fits}

The available lines in the fitting procedure were H$\beta$, {[}OIII{]}, {[}OI{]}, {[}NII{]}, H$\alpha$ and {[}SII{]}, however this paper will focus on the H$\alpha$ ($\lambda 6563$) emission, fitted alongside the {[}NII{] $(\lambda 6548, 6583$)} lines. A following paper will cover the full range of emission lines. The line fitting was carried out on a cube smoothed spatially using the mean value within a 3x3 kernel.

The redshift used for fitting the lines in \textsc{kubeviz} was set at $z=0.045$, calculated by \textsc{kubeviz} using the spectral position of the H$\alpha$ line at the central point of the galactic disk.

As the noise values in the `stat' datacube produced by the MUSE pipeline underestimate the total error, the option was selected in \textsc{kubeviz} to compute the nominal standard deviations of the measurements after renormalising the covariance matrix assuming a reduced $\chi^2 = 1$ \citep{2016MNRAS.455.2028F}.
The continuum under each line was calculated between 80 and 200$\AA$ from each line, omitting regions containing other known emission lines and using values within the 40th and 60th percentiles. An initial value of $20$km~s$^{-1}$ was set for the narrow linewidth to prevent \textsc{kubeviz} returning an error when fitting regions with low velocity dispersion, in particular the upper region of the galactic disk and the central region of the star forming knot to the NE of the field.

Figure \ref{figure:snJO201} shows the $H\alpha$ S/N map resulting from the line fits. We rejected fits with S/N $<$ 3, as well as unrealistic fits flagged by \textsc{kubeviz}, which have zero velocity or velocity error.

For each fitted spaxel the fitting process yielded a central wavelength, intensity, and width of the H${\alpha}$ line, that could in turn be converted into velocity, flux and velocity dispersion. 

\begin{figure}\centering
\includegraphics[width=0.5\textwidth]{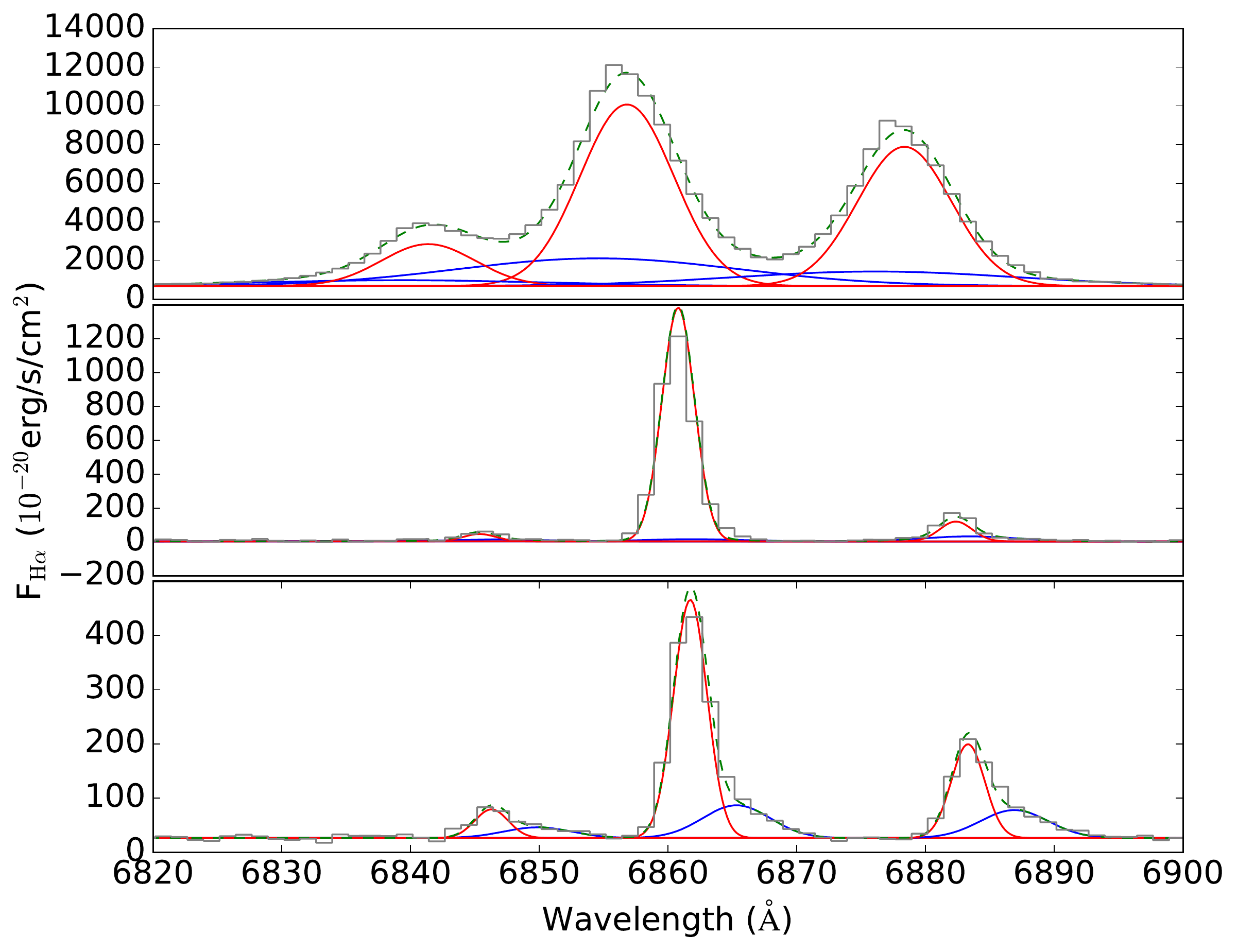}
\caption{The observed H$\alpha$, NII emission line group in 3 different spaxels within JO201 (black histograms) together with a 2 component fit (red and blue curves correspond to individual component and the green dashed line their sum). \textit{Top panel}: The central region of the galaxy, broadened by the AGN. \textit{Middle panel}: A region within the north-eastern star-forming blob.
\textit{Bottom panel}: A region in the outer part of the disk displaying obvious tails redward of the emission.
}\label{figure:spectrazoom}
\end{figure}

We visually inspected the emission-line fits and found that in most cases the fits describe the line profiles well. In some regions, however, the line profiles are  more complex and thus multi-component fits were needed. Three example spectra extracted from different spaxels within the galaxy are shown in Figure~\ref{figure:spectrazoom}. The top panel shows the central region of the galaxy where an AGN has clearly broadened the lines. In this case, two gaussian components (red and blue lines) fit the emission lines better than a single component. The middle panel shows a star-forming region (large H$\alpha$ blob with high S/N NE of the galaxy disk, visible in Figure~\ref{figure:snJO201}) well fitted by a single gaussian. In the bottom panel instead, we see non-gaussian emission (with one-sided wings) from a region in the outer part of the disk. As shown in Section~\ref{sec:2comp}, we find many of these non-gaussian emission lines in the galaxy. In some (more extreme) cases, two separate peaks can be distinguished (see bottom panels of Figure~\ref{figure:lineprofiles}). As will be established and further discussed in the following sections, the shapes of the lines in these cases are indicative of gas trailing behind the disk of the galaxy, or even multiple physical blobs present in the same line of sight.


\subsection{Double Component Fits} \label{sec:2comp}

Owing to the variety and complexity of the observed emission line profiles, in this section we present the two-component fits to the H$\alpha$ emission across the galaxy. In the following sections, we will combine the two-component fits with single-component fits to construct maps of the emission in JO201.

We re-fitted all of the emission lines, using \textsc{kubeviz}, setting two gaussian components to prioritise convergence on narrow and broad components of the emission, based on the value of their velocity dispersion.
The resulting fits yield line centers ($\lambda_\mathrm{narrow}$ and $\lambda_\mathrm{broad}$), fluxes ($\mathrm{F}_\mathrm{narrow}$ and $\mathrm{F}_\mathrm{broad}$), and velocity dispersions ($\sigma_\mathrm{narrow}$ and $\sigma_\mathrm{broad}$) for each spaxel. From these parameters we could then investigate in each case which component was the brightest or bluest and reveal which regions exhibited strong wings in the emission lines or double-peaking. As before, we rejected fits with S/N $<$ 3 (in either component) as well as those rejected by \textsc{kubeviz}.

For each fitted emission-line, flags were allocated based on the following parameters.

\begin{itemize}
\item 0 for spectra in which the narrow component is redward of the broad component. I.e.  $ \lambda_\mathrm{narrow} > \lambda_\mathrm{broad}$.

\item +1 for spectra in which the narrow component is blueward of the broad component:  $\lambda_\mathrm{narrow} < \lambda_\mathrm{broad}$.

\item +2 where the narrow component is brighter than the broad component: $\mathrm{F}_\mathrm{narrow} > \mathrm{F}_\mathrm{broad}$.

\item +4 where the separation between the two lines is greater than the sum in quadrature of their sigmas, Suggesting that two distinct lines are present:  $\lambda_\mathrm{narrow}-\lambda_\mathrm{broad}>\left({\sigma_\mathrm{narrow}^2 + \sigma_\mathrm{broad}^2}\right)^{1/2}$.

\end{itemize}
For example, a region flagged as 5 will be comprised of flags +1 and +4, i.e. the narrow component will be blueward of and fainter than the broad component, with both components being significantly separated from each other. 

\begin{figure}\centering
\includegraphics[width=0.5\textwidth]{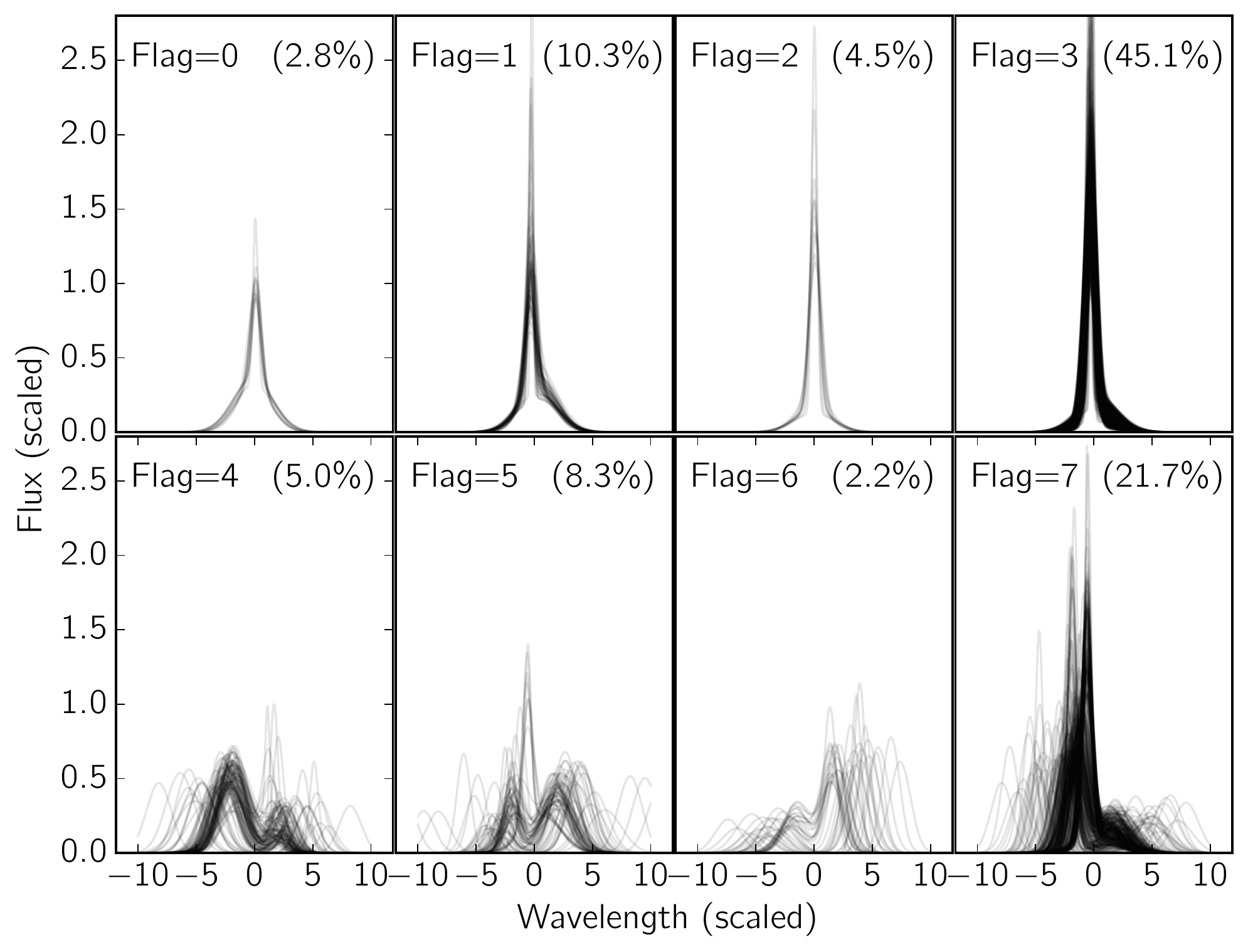}
\caption{Profiles of double-component fits made to H$\alpha$ emission lines in JO201, separated into the combined flags described in Section~\ref{sec:2comp}. Each of the plotted profiles has been normalised in position width and height to facilitate the comparison of the shape of the profiles. Percentages show the number of spaxels with each flag value out of the total number of spaxels fitted with 2 components. The flag values are mapped in the context of the galaxy in Figure~\ref{figure:comp_map}.} \label{figure:lineprofiles}
\end{figure}

Our flagging system allows the separation of different line profiles, as shown in Figure~\ref{figure:lineprofiles}, where all the double-component line fits are split into the different flags. For this plot the fits were normalised to a standard velocity, width and height to facilitate their comparison. 

\begin{figure}
\includegraphics[width=0.45\textwidth]{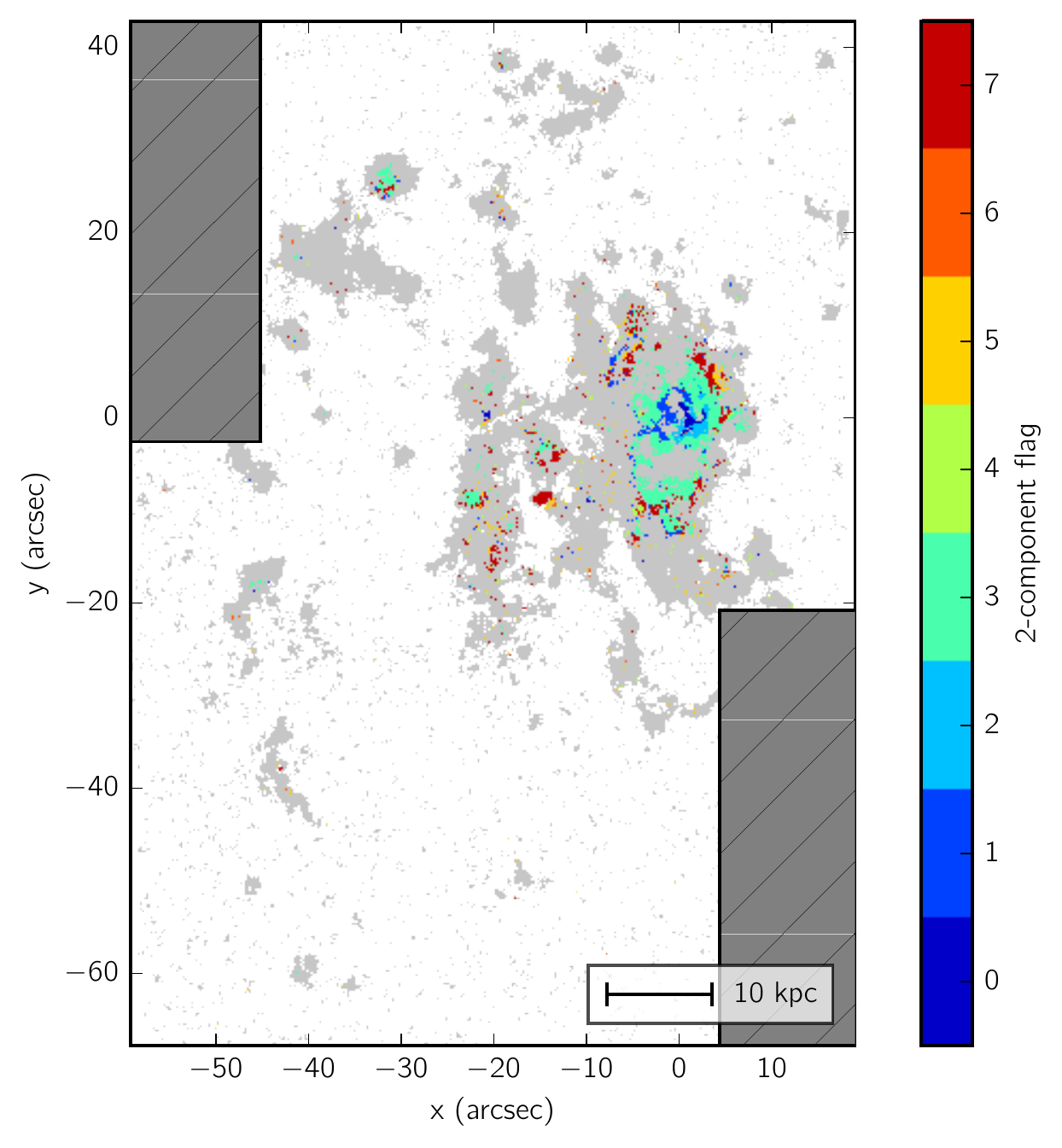}
\caption{Map of fitted spaxels in JO201, separated into regions where double-component fits were used (coloured) or where single-component fits were sufficient to fit the H$\alpha$ emission (grey). We chose the double-component fits when both the narrow and broad component were detected to greater than $3\sigma$, otherwise a single component (also limited to $3\sigma$) was used. Colours other than grey correspond to the flags attributed to each 2-component fit (c.f. Figure~\ref{figure:lineprofiles}), as indicated by the colourbar.}\label{figure:comp_map}
\end{figure}

It is interesting to map the location of the different line profiles in the galaxy. This is shown by the coloured regions in Figure~\ref{figure:comp_map}. 

Emission-lines with Flag~$<3$  are predominant in the centre of the galaxy. These fits  correspond to cases where the 2 components are aligned or not significantly misaligned (see top row in Figure~\ref{figure:lineprofiles}). 
The cause of these types of line profiles is the presence of an AGN. The active nucleus not only causes line broadening (Flag~$=0$), but, in some cases, it can generate outflows visible as mildly visible wings to one side of the broadened lines (e.g. Flag~$=$1). Lines with Flags~$=$0, 1 and 2 account for 18\% of the 2-component fits with $S/N>3$. 

Lines with Flag~$=3$ instead show skewed emission lines, exhibiting wings on the redward sides of bright emission lines. These trailing wing profiles are common (45\%), and are typically found in the disk. 

Fits with Flag~$\geq 4$ are more extreme cases where there is a significantly displaced second component (37\% of all 2-component fits, see bottom row of Figure~\ref{figure:lineprofiles}). 
These types of fits were the best description of the emission lines in the regions around the edges of the galactic disk as well as the diffuse regions around the stripped tails.
Their location and morphology are consistent with the scenario in which JO201 is primarily undergoing gas stripping along the line of sight, with the stripped gas being dragged behind the galaxy by ram pressure.
The stripped material could be arranged either in the form of a physical tail behind the galaxy, or separate blobs along the same line of sight. 
This interpretation is in agreement with the environmental  analysis presented in Section~\ref{sec:env}, and further supported by the kinematic analysis of the stellar and gaseous components of the galaxy, presented in Section~\ref{sec:kinematics}.

\subsection{Final fits} \label{sec:1vs2comp} 

To identify the best fit in each fitted spaxel we compare the results of the single versus double-component fits in each case.
We use the $S/N$ ratio of the single-component fit and the narrow and broad components of the double-component fits for this purpose.
 
We use the double-component fits when both the narrow and broad component are detected to greater than $3\sigma$, otherwise a single component (also limited to $3\sigma$) is used. Figure \ref{figure:comp_map} shows the regions where double-component (colour) versus single-component (grey) fits are used. The figure shows that although single-component fits characterise well the emission in most parts of the galaxy, two-component fits are needed in the central region and in some parts of the disk and stripped tails. 

Unless otherwise stated, the above mentioned combination of single and double-component fits is used in the rest of this analysis.

\subsection{$\mathrm{H}\alpha$ distribution}

\begin{figure}\centering

\includegraphics[width=0.45\textwidth]{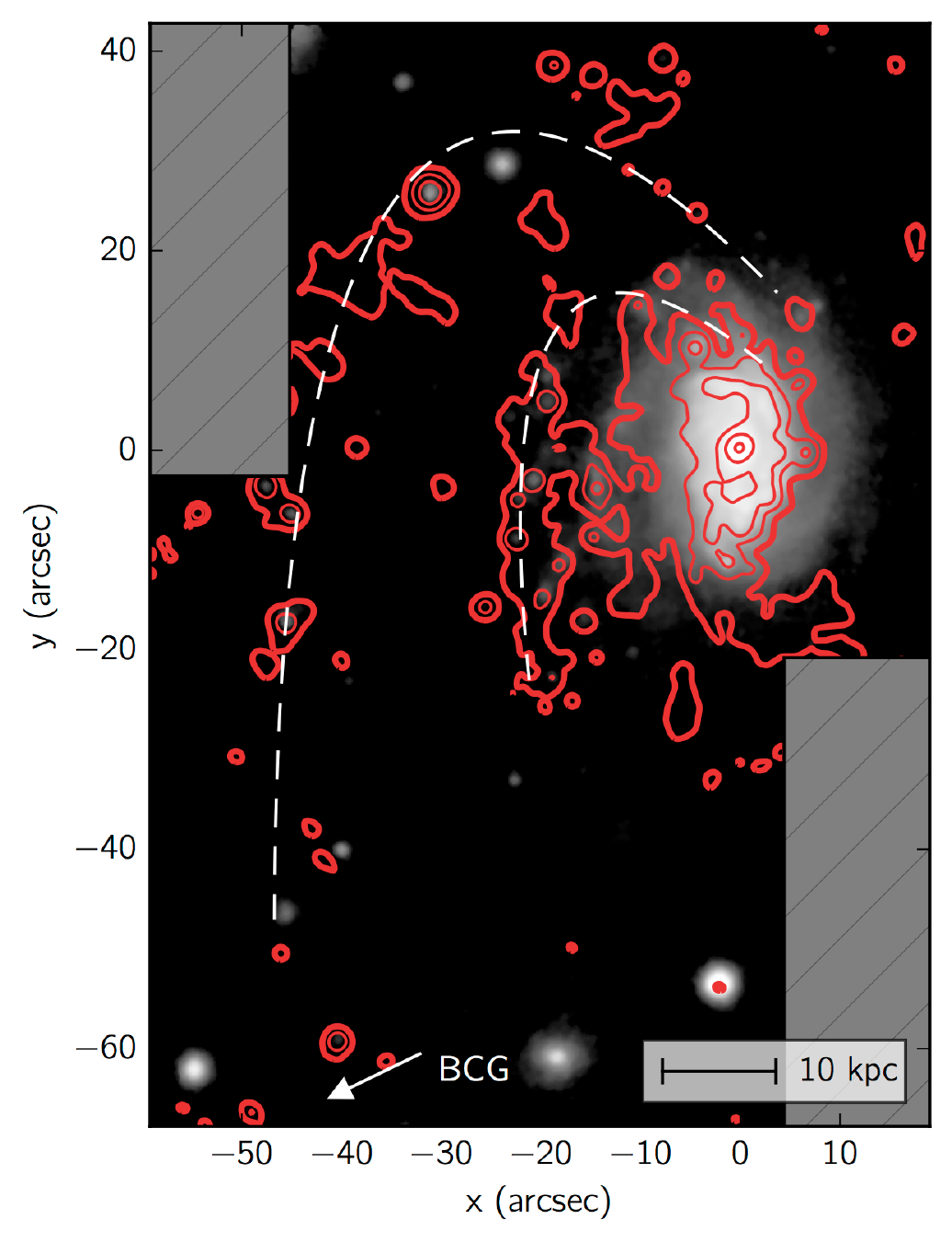}
\caption{Contours of H$\alpha$ ($\lambda 6562.82$\AA) overlaid on continuum fit around H$\alpha$ line. Steeper contours on the western edge of the disk suggest increased star formation activity resulting from the interaction with the ICM. Contours correspond to surface brightness levels of $6.67\times 10^{-19}$, $3.55\times 10^{-18}$, $1.89\times 10^{-17}$, $1.00\times 10^{-16}$, $5.33\times 10^{-16}$ $\mathrm{erg}\, \mathrm{s}^{-1} \mathrm{cm}^{-2}$ from low to high. The pattern of the stripped material appears to follow the shape of the spiral arms (see dashed white curves as reference) as discussed in the text. The white arrow indicates the direction toward the BCG.}\label{figure:linemaps}
\end{figure}

The spatial distribution of the fitted line flux in $\mathrm{H}\alpha$ is shown as red contours overlaid on the continuum fit from \textsc{kubeviz} in Figure~\ref{figure:linemaps}. It is visible in the figure that the emission from the H$\alpha$ component shows a large offset, down to a surface brightness of ${\sim}0.7 \times 10^{-18} \mathrm{erg}\, \mathrm{s}^{-1} \mathrm{cm}^{-2}$, from the disk of the galaxy due to the effect of RPS. Furthermore, the bright ridge of $\mathrm{H}{\alpha}$ on the west side of the galaxy may evidence the interaction with the hot ICM compressing the gas and forming a bow shock, increasing the star formation activity and thus the H$\alpha$ emission. To the East of the galaxy disk (roughly in the direction towards the BCG, as shown by the arrow in the figure) the H$\alpha$ emission is drawn out into ``tentacles" of trailing gas which connect brighter knots. These knots are likely to be previously stripped gas which has collapsed under gravity to undergo intense star formation.
The pattern of the diffuse emission and the collection of brighter knots observed in the plot suggest an unwinding of the spiral arm structure (see white dashed lines in Figure~\ref{figure:linemaps}). 
A possible explanation is that strong ram-pressure ``wind" can cause the spiral arms to unwind.
This effect has not been observed in hydrodynamical simulations (yet) due to the difficulty of reproducing realistic spiral arms. 
Observationally the effect has not been reported either, although studies of jellyfish galaxies have mostly focused on stripping of galaxies in the plane of the sky, where the tails are more easily detected. The proposed effect, if true, would be better appreciated in galaxies like JO201, that are experiencing intense face-on stripping and are viewed along the direction of the stripping (i.e. along the line of sight). 
We plan to utilise simulations to address this idea in future studies.

Although the jellyfish tails in JO201 extend to ${\sim} 50$~kpc from the disk, it is likely that they are much larger given the strong projection effects expected for this galaxy and the fact that the stripped material covers essentially the entire field of view of the MUSE mosaic.

Our measurements are consistent with recent observations of other jellyfish galaxies in clusters. For example, from   
deep H$\alpha+$[NII] wide-field imaging with 
MegaCam at the CFHT, \citet{Boselli2016} found tails of diffuse ionised gas in NGC 4590 extending $\sim80$~kpc from the disk in projection (at a surface brightness limit of a few 10$^{-18}$ erg~s$^{-1}$~cm$^{-2}$).
Using MUSE at the VLT, \citet{Fumagalli2014} observed the jellyfish galaxy ESO137-001 in the Norma cluster, and found that at a surface brightness $\sim$10$^{-18}$ erg~s$^{-1}$~cm$^{-2}$, the tails  extend to $>30$~kpc from the disk. 
These examples are quite different from JO201 and its environment: NGC 4590 is a very massive late-type galaxy falling into a low-mass cluster,  ESO137-001 is a lower-mass spiral falling in a very high mass cluster, and JO201 is a massive spiral falling into a very massive cluster. Although the exact comparison of the tail extent is subject to projection effects and depth of the observations, we can confidently conclude that the observed tails in all cases clearly extend well beyond the galaxy (out to at least $30-80$~kpc), which implies that RPS has a wide range of action.

\subsection{3D Rendering $\mathrm{H}\alpha$ Emission}

To visualise the distribution of the $\mathrm{H}\alpha$ gas, a 3D model was produced by combining projected space and velocity space. 

\begin{figure}\centering
\includegraphics[width=0.4\textwidth]{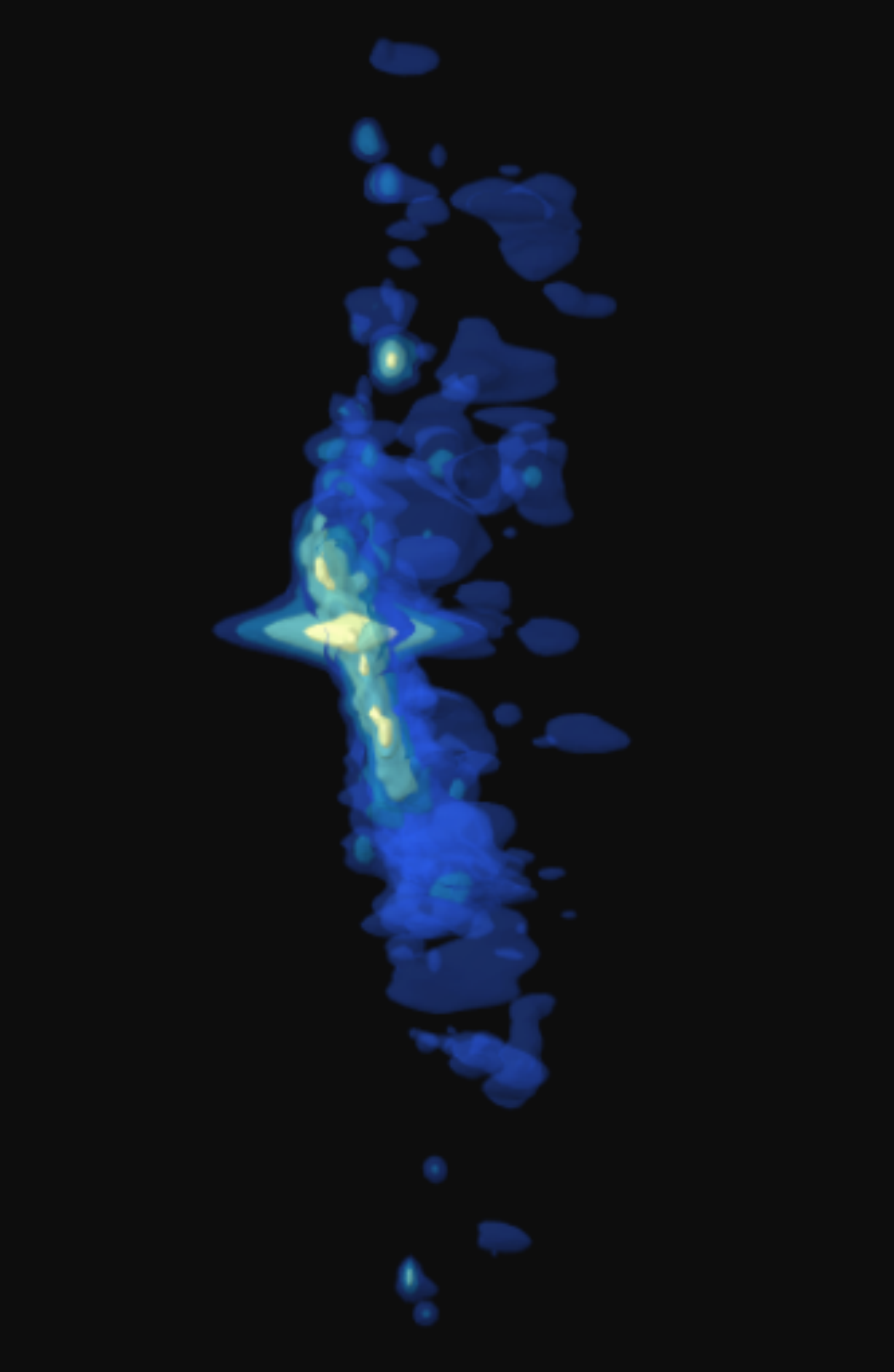}
\caption{3D visualisation of $\mathrm{H}\alpha$ gas extracted from \textsc{kubeviz} fit produced using the method presented in \citet{Vogt2016} and \citet{Vogt2016b}. X and Y axes show projected distance, with the Z axis representing the wavelength/velocity and linewidth/sigma. Surface contours are plotted with different colours, from yellow to dark blue. The units and contours are arbitrary as they have been chosen at levels which highlight interesting features in the $\mathrm{H}\alpha$ gas, such as the bright central emission, knots of $\mathrm{H}\alpha$ and the extended diffuse gas. An interactive version of this figure is accessible online on the HTML version of this paper, and at \url{web.oapd.inaf.it/gasp/publications.html}.}\label{figure:3dmodel}
\end{figure}

The fitted $\mathrm{H}\alpha$ line was used to produce a ``cleaned'' cube comprised of only $\mathrm{H}\alpha$ emission, onto which surface contours were wrapped using the velocity distribution in place of the spatial distribution of the gas along the line of sight.
We selected contour levels manually to optimally separate different points of interest, such as the bright central $\mathrm{H}\alpha$ emission, the knots within the tails and the diffuse extended gas. 
The resulting 3D  model is presented in Figure~\ref{figure:3dmodel}. An interactive version of the figure is available on the online version of this paper, and at \url{web.oapd.inaf.it/gasp/publications.html}.

Viewed face-on, the 3D model visualises the distribution of the $\mathrm{H}\alpha$ gas and highlights the bright knots within the tails (like Figure~\ref{figure:linemaps}). If the cube is rotated to get a view through the y-axis, the rotation of the gas is visible both within the disk and in the tails, along with a dispersion-dominated nuclei. From this perspective (and also when viewing through the x axis), the diffuse component of the gas can be seen at higher line-of-sight velocities than the denser knots.

\section{Kinematics}\label{sec:kinematics}

\subsection{Absorption Line Kinematics}\label{sec:ppxffit}

\begin{figure*}\centering
\includegraphics[width=0.9\textwidth]{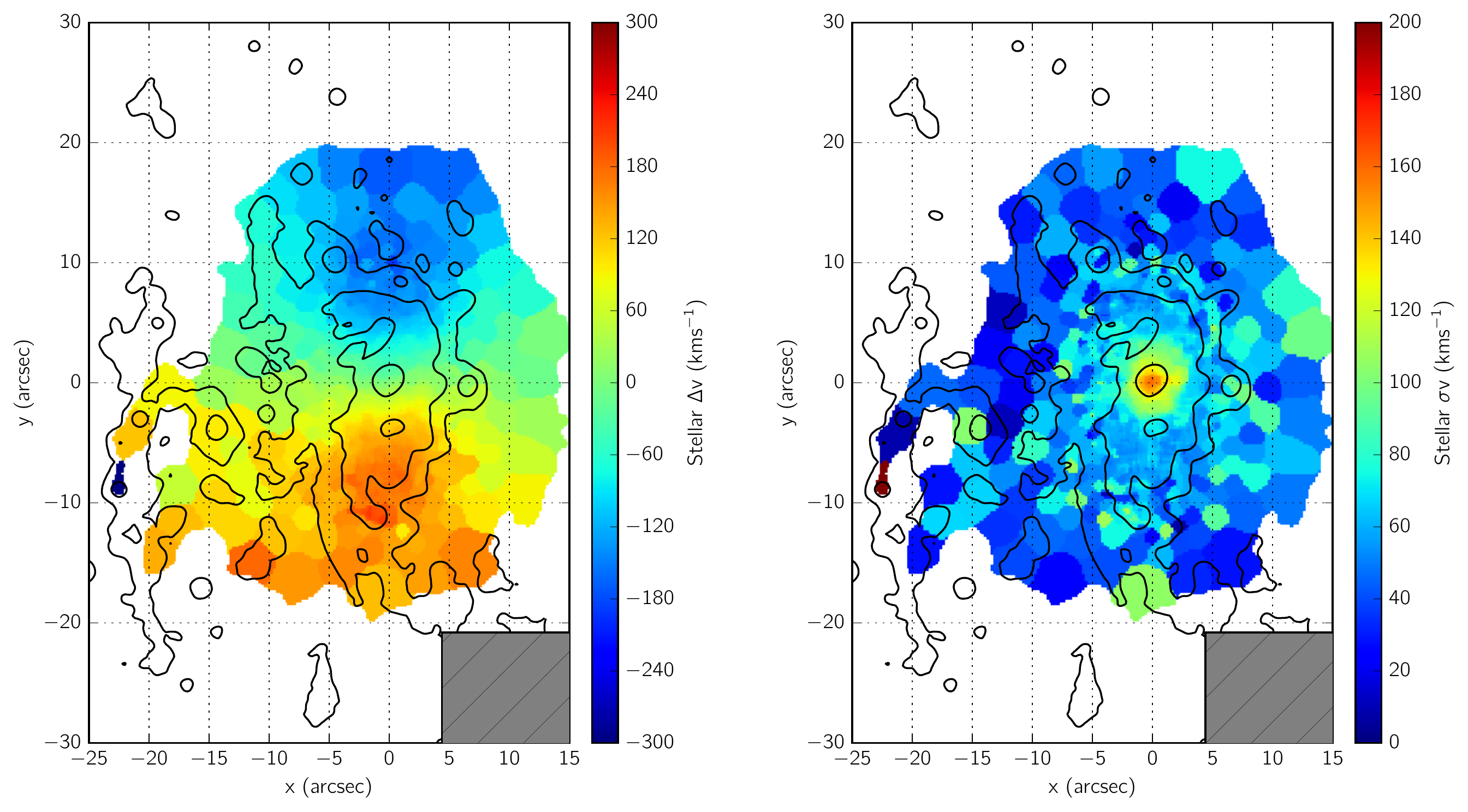}
\caption{Stellar kinematics (\textit{left:} line-of-sight velocity, \textit{right:} velocity dispersion) derived from ppxf fits binned using a Voronoi Tessellation to obtain a S/N of 15. H$\alpha$ contours are shown overlaid. As expected for the case of RPS, the stellar component shows smooth, generally unperturbed kinematics. This is consistent with a hydrodynamical interaction as opposed to a gravitational one.}\label{figure:stellarkin_map}
\end{figure*}

\begin{figure*}
\includegraphics[width=0.5\textwidth]{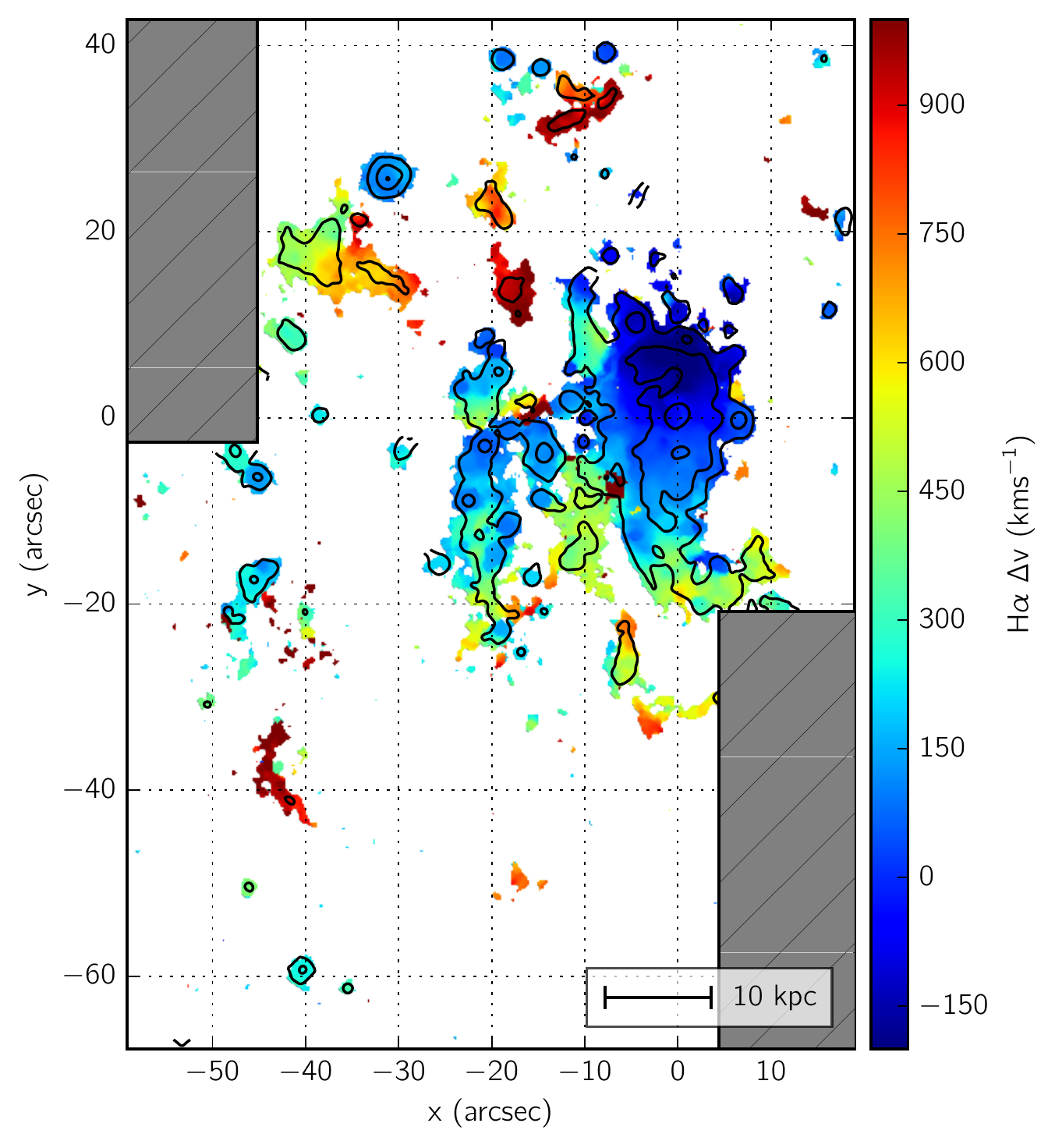}
\includegraphics[width=0.48\textwidth]{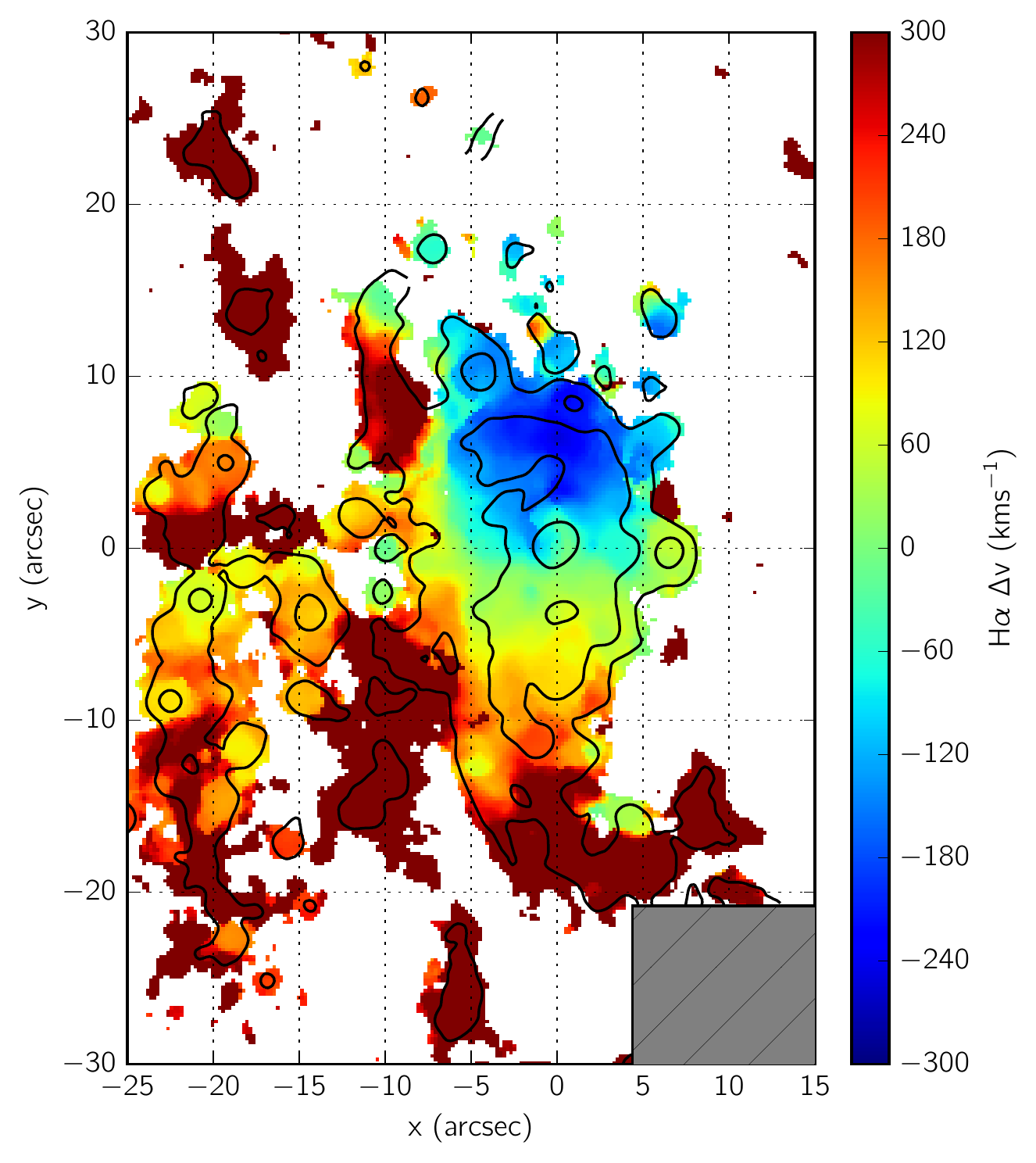}
\caption{\textit{Left:} Velocity map produced of H$\alpha$ ($\lambda 6562.82$\AA) emission relative to the galaxy systemic velocity comprised of a combination of single component and two-component fits as described in section \ref{sec:2comp}. The velocity map has been convolved with a median filter with a $3 \times 3$ pixel kernel for visualisation purposes, in addition to the data being smoothed using a mean filter with a $3 \times 3$ kernel before fitting. \textit{Right:} Velocity map as on left, cropped to show the disk region with a rescaled colour map to better exhibit the H$\alpha$ kinematics of the disk region. Black contours on both figures indicate the $\mathrm{H}\alpha$ flux for context.}\label{figure:halpha_velmap}
\end{figure*}

In order to extract the stellar kinematics, we extracted the absorption line properties from the galactic extinction corrected cube using the Penalized Pixel-Fitting (pPXF) code \citep{Cappellari2004}, that fits the observed spectra with a set of templates. We spatially binned the spectra, using a Voronoi tessellation dependent on the signal to noise of our spectra (S/N=15), as described in \citet{Cappellari2003}. In particular, we used a Weighted Voronoi Tessellation \citep{Diehl2006} in order to take care of possible significant gradient in the S/N.
For our analysis we used the stellar population templates by \citet{Vazdekis2010}, i.e. a set of SSP spanning a range
in metallicity (6 different metallicities from $[M/H]=-1.71$ to
$[M/H]=0.22$) and age (26 ages from 1 to 17.78 Gyr) calculated with a
Salpeter IMF \citep{Salpeter1955} with a slope of 1.30 and the Padova 1994 isochrones \citep{Bertelli1994, Girardi1996}. We excluded from the fit the red part of the spectra, where the theoretical stellar libraries
have a poor resolution and the observed spectra are strongly
contaminated by the sky lines.  
Spurious sources (foreground stars, background galaxies) have been
masked before performing the fit. 
We derived the rotational
velocity, the velocity dispersion and the two h3 and h4 moments.  This
allows us to derive for each Voronoi bin a redshift estimate, that will
then be used as input for the stellar population analysis.

The stellar kinematics maps (not corrected for the inclination of the galaxy) are shown in Figure~\ref{figure:stellarkin_map} with the velocity on the left panel and the velocity dispersion on the right. The velocity map shows that the stellar component follows a uniform  rotation, with maximum velocity $\simeq 155 $km~s$^{-1}$.  The velocity dispersion profile of this galaxy is also smooth, with increasing values towards the centre that peak at $\simeq 150 $km~s$^{-1}$. 

To the south-east of the disk, two parallel tails are visible. 
It is unlikely that these tails arise from tidal interactions, since i) they are seen only in one side of the galaxy, ii) spectrophotometric modelling shows they have very young stellar populations (Bellhouse et al. in preparation), and iii) they follow the main rotation of the galaxy. As will be discussed in Bellhouse et al. (in preparation), these young stellar tails most likely formed in-situ from stripped gas. 

 Overall, the stellar kinematics of JO201 is consistent with that of a gravitationally undisturbed galaxy. 

\subsection{Emission Line Kinematics}\label{sec:emkin}

In order to produce maps of the kinematics of the gaseous component in JO201, we extracted the velocity and velocity dispersion of the H$\alpha$ ($\lambda 6562.82$\AA) line from the line fits presented in Section~\ref{sec:linefits}.

\begin{figure}\includegraphics[width=0.5\textwidth]{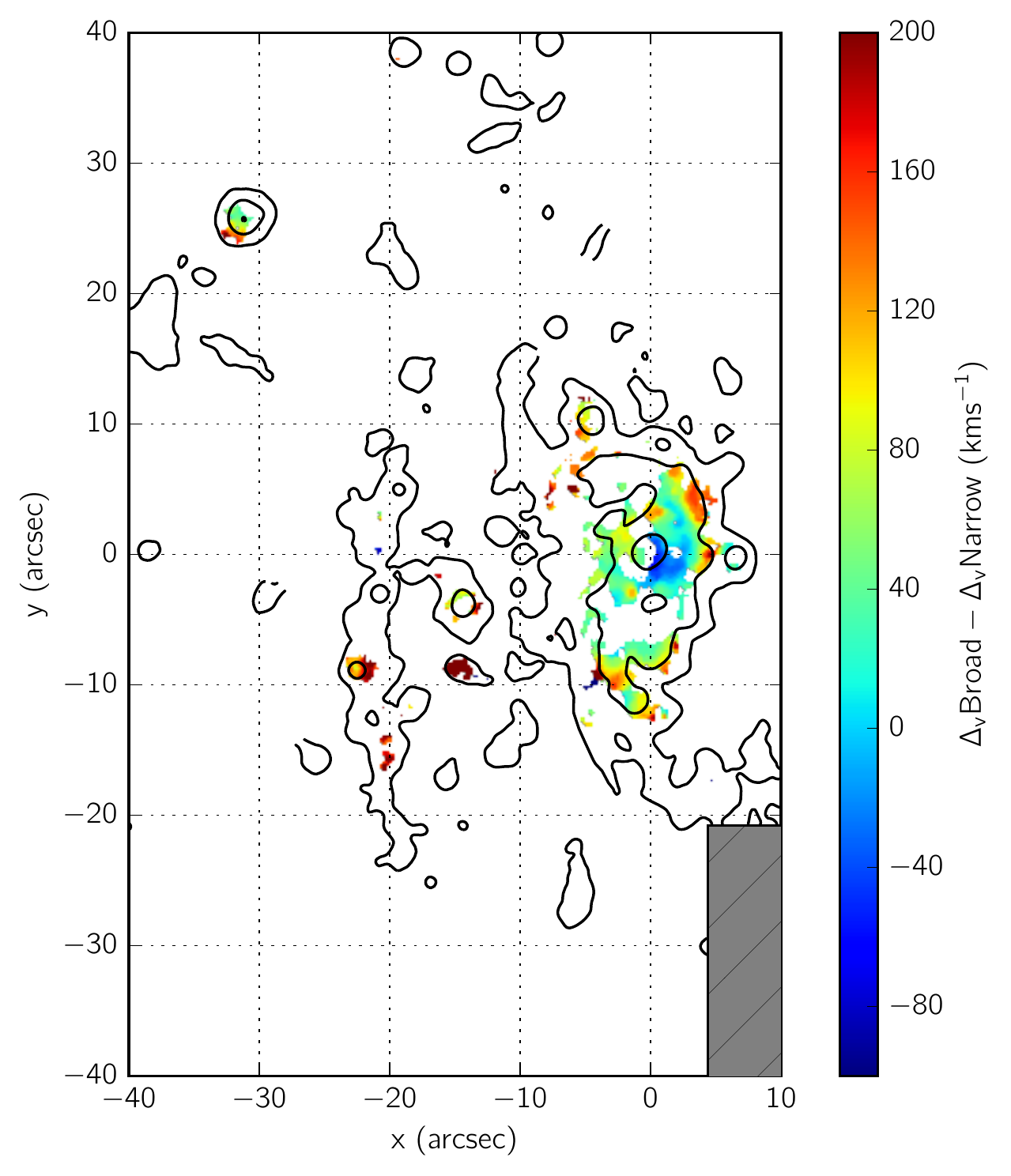}
\caption{Velocity offset between broad and narrow components of H$\alpha$ ($\lambda 6562.82$\AA) line where double component fit is available. The region of blueward broad component indicating a tail oriented towards the observer is likely to originate from an outflow associated with the central AGN. The redward broad component in the outskirts of the disk evidence a tail in the velocity distribution produced by increased stripping intensity at the outer edges of the disk. Black contours indicate the $\mathrm{H}\alpha$ flux for context}\label{figure:veldiff}
\end{figure}

\begin{figure}
\includegraphics[width=0.5\textwidth]{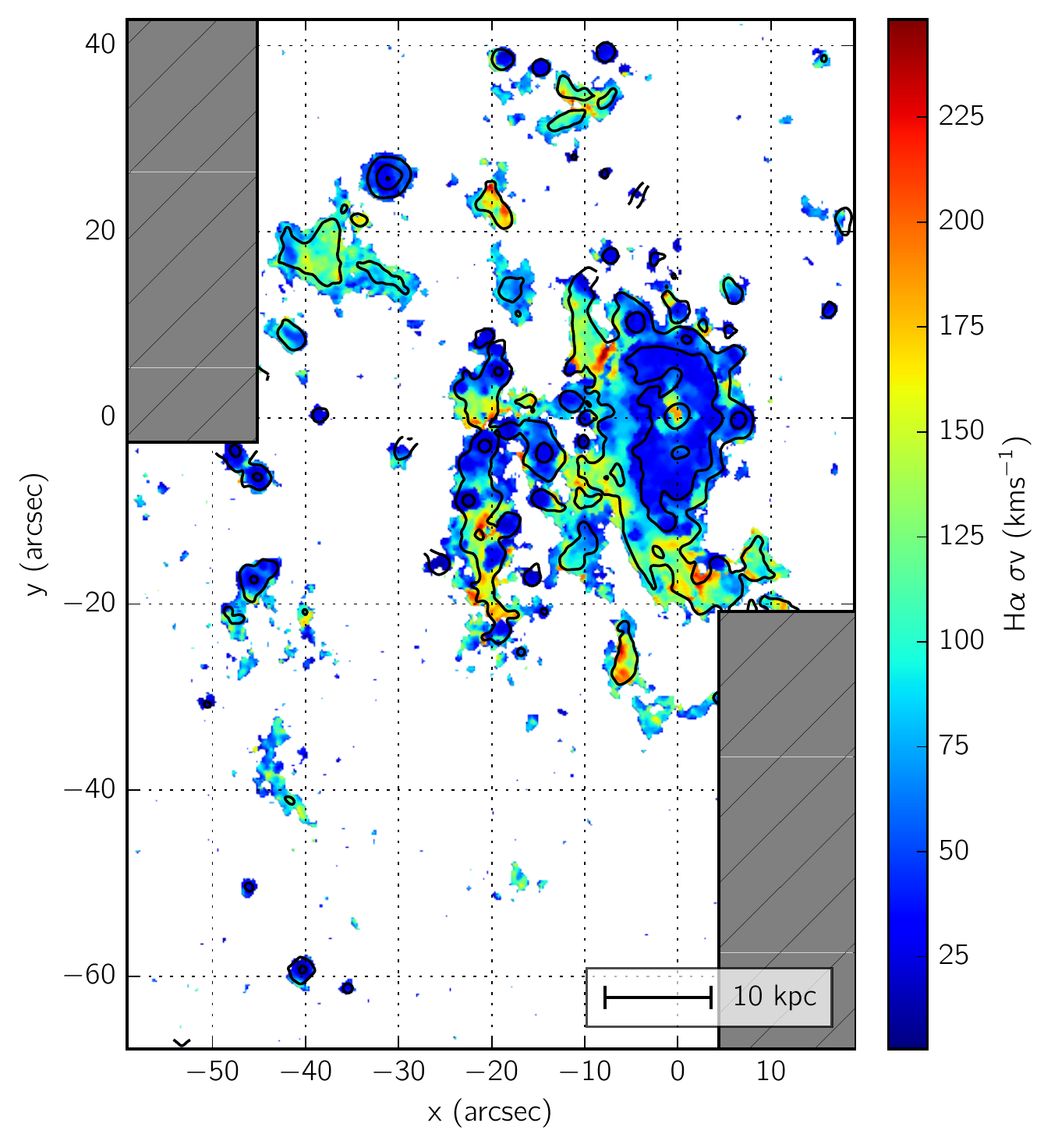}
\caption{Velocity dispersion map produced using H$\alpha$ ($\lambda 6562.82$\AA) emission from combined single component and two-component fits as described in section \ref{sec:2comp}, corrected for instrument resolution by \textsc{kubeviz}. The velocity dispersion map has been convolved with a median filter with a $3 \times 3$ pixel kernel for visualisation purposes.}\label{figure:halpha_veldisp}
\end{figure}

\begin{figure}
\includegraphics[width=0.5\textwidth]{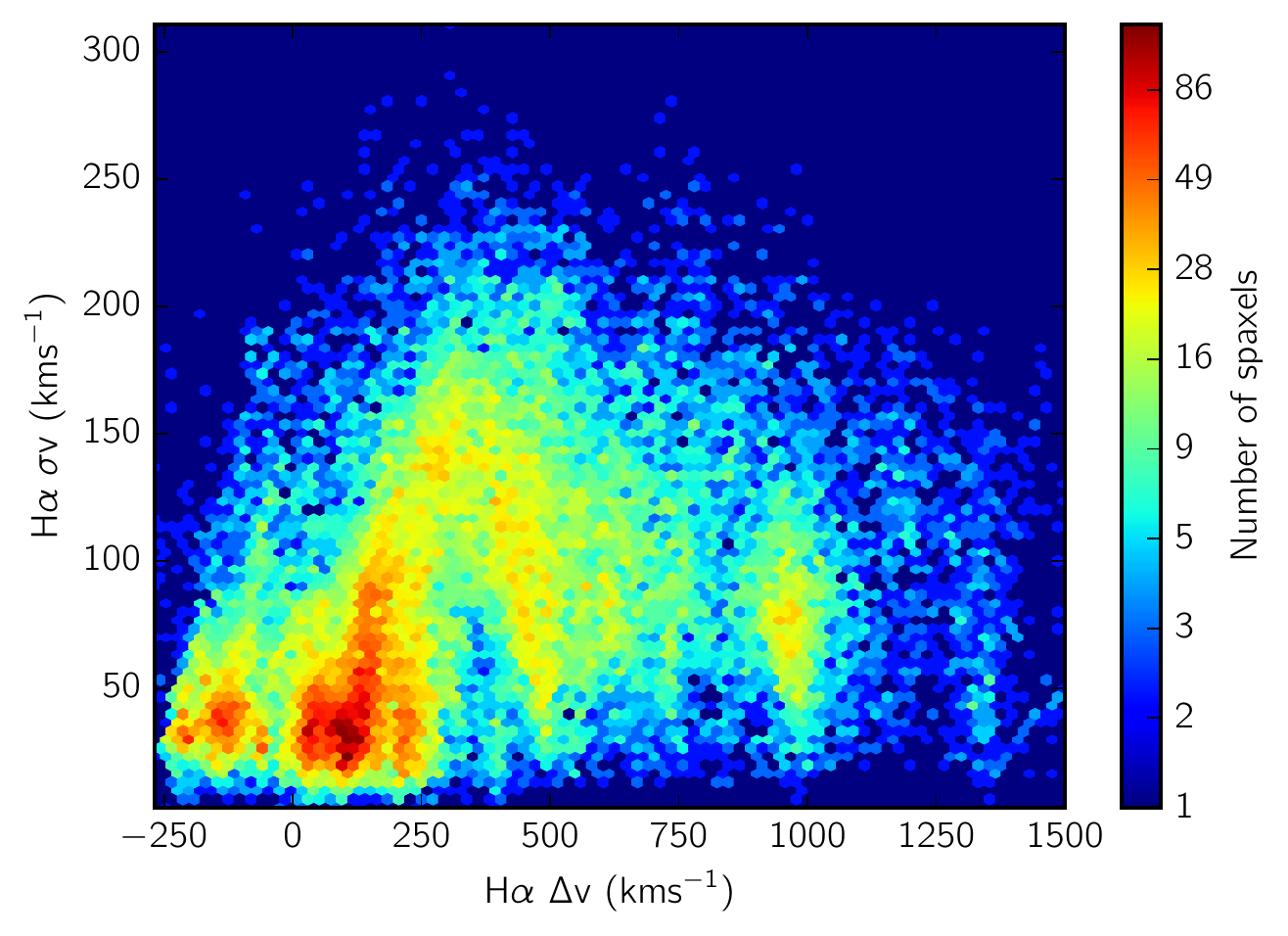}
\caption{Distribution of spaxels arranged by velocity on x axis and sigma on y axis. The plot shows a bimodal distribution around $0$km~s$^{-1}$ from the rotation curve of the galaxy, followed by an increase in dispersion with velocity due to the onset of turbulent motions within the stripped gas. This trend reverses above $350$km~s$^{-1}$ as the dispersion falls back down, likely to be caused by the recollapse of cooling stripped gas causing the dispersion to decrease.}\label{figure:kinematics_hist}
\end{figure}

The left panel of Figure~\ref{figure:halpha_velmap} shows a map of the line-of-sight velocities of the $\mathrm{H}\alpha$ emission (not corrected for inclination). A zoomed version focusing only on the disk (with a smaller range of velocities) is shown on the right panel of the figure.
Overall, the gas component displays a clear rotation within the disk.  
The rotational velocity gradient of the galaxy is somewhat retained in the stripped material to at least $50\mathrm{kpc}$ in projected distance from the disk, although beyond the central disk, the stripped material outside the galaxy's main body is notably residing at higher line-of-sight velocities (up to ${\sim}900$km~s$^{-1}$) with respect to the disk. 
This highly redshifted emission reflects the intense RPS acting on the galaxy along the line of sight. As the galaxy  plunges towards the observer, the gas stripped from the (less shielded) outskirts of the galaxy is dragged away from the galaxy.
As already mentioned, this interpretation is in agreement with the extreme line-of-sight velocity of JO201 with respect to the cluster (Section~\ref{sec:env}), which suggests that the velocity vector of the galaxy is mostly in the line of sight and thus the majority of the stripped material must reside behind the galaxy. 

Inside the disk, the maximum observed rotational velocity of the ionised gas toward the observer is ${\sim} 144$km~s$^{-1}$, extending beyond ${\sim} 179$km~s$^{-1}$ away from the observer at the southern end of the disk, up to where gas starts to be visibly stripped, both along the plane of the sky and in the redshift direction. From a quick comparison of the rotation of the gas (Figure~\ref{figure:halpha_velmap}, right) and the stars (Figure~\ref{figure:stellarkin_map}, left), it is noticeable that the rotational velocity of the gas is significantly higher than that of the stars. A careful comparison of the stellar and gas rotation curves is presented in Section~\ref{sec:RC}.

We further compare the kinematics of the broad and narrow components of the H$\alpha$ line in Figure~\ref{figure:veldiff} for those regions where a double-component fit was available. The colours in the figure highlight the difference between the velocities of the 2 components. The central region showing a blueward broad component indicates a tail oriented towards the observer. This is likely to originate from an outflow associated with the central AGN. The redward broad component in the outskirts of the disk is suggestive of tails produced by increased stripping intensity at the outer edges of the disk, consistent with outside-inward stripping. 

The line-of-sight velocity dispersion of the gas component is shown in  Figure~\ref{figure:halpha_veldisp}. The figure shows that the velocity dispersion in the central region of the disk is $150-180$km~s$^{-1}$. The lowest velocity dispersions are observed in the dense knots with large H$\alpha$ fluxes, with typical values ranging around $20-40$km~s$^{-1}$. The stripped diffuse gas bridging between star-forming knots appears to have a higher velocity dispersion, with values around $140$km~s$^{-1}$. 

\begin{figure*}\centering
\includegraphics[width=\textwidth]{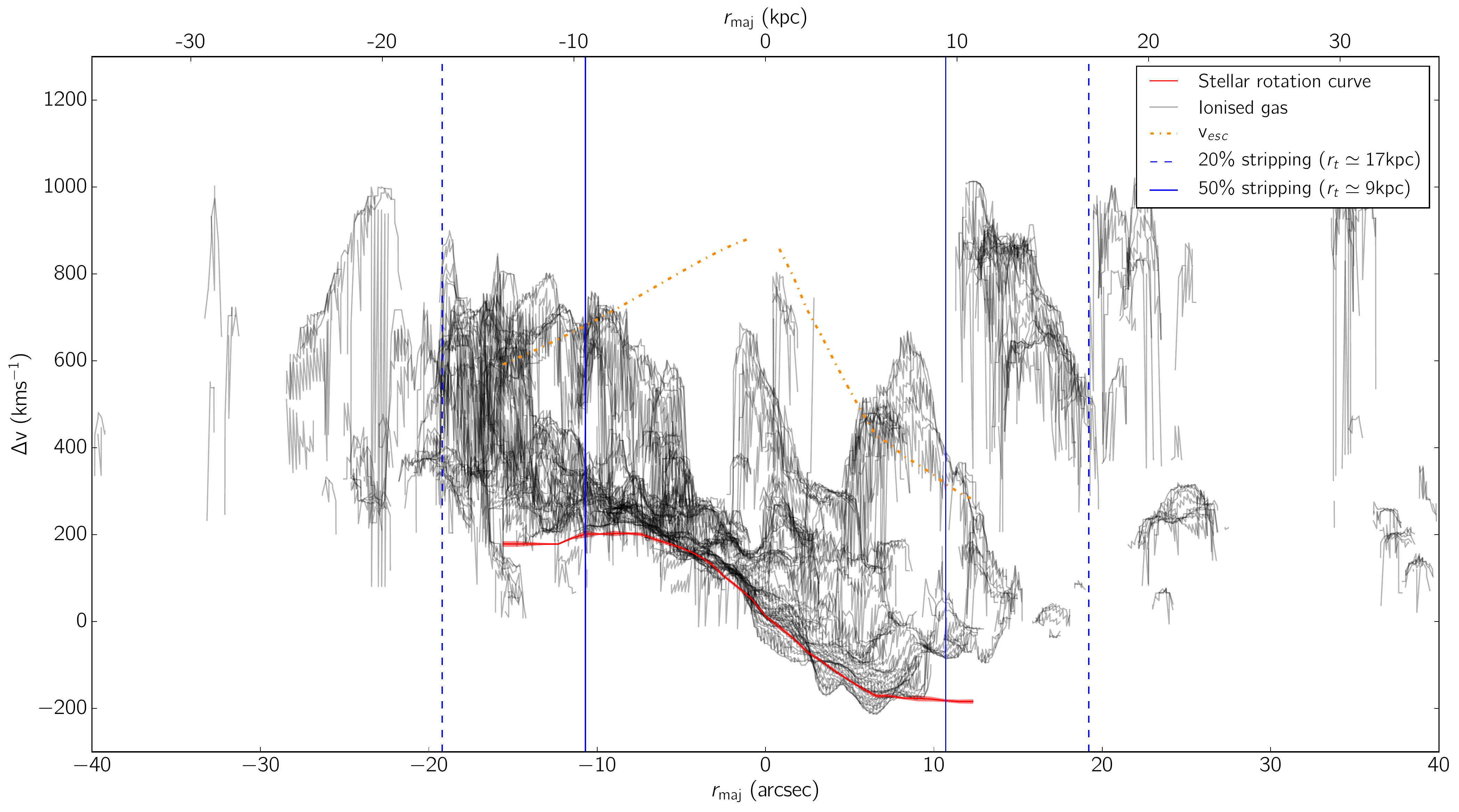}
\caption{The rotation curve along the kinematic major axis $\mathrm{r}_\mathrm{maj}$ for the stellar component of JO201 is plotted in red. The solid black lines show the velocity of the ionised gas along strips parallel to the direction of $\mathrm{r}_\mathrm{maj}$. The orange dot-dashed line shows the escape velocity calculated as described in the text. The vertical blue dashed and solid lines further show the truncation radii at which 20\% and 50\% of the total gas mass is expected to have been stripped respectively, as described in Section~\ref{sec:env} (see lines in  Figure~\ref{figure:Env2}). The gas stripping is made clear by the increased line-of-sight velocity of the ionised gas to the left and right of the disk caused by regions of stripped gas being slowed by the ICM with respect to the infalling galaxy. Rotating clumps of gas can also be seen at higher $\Delta v$ in the stripped gas, rotating around their own centres, away from the motion of the galaxy. 
It is likely that much of this gas has become unbound from the galaxy due to its high velocity offset, particularly where it approaches and passes the indicated escape velocity, however the gas retains the angular momentum of its previous rotation within the galaxy. The region of gas which remains mostly bound to the disk is best illustrated by the dashed lines in Figure~\ref{figure:starsgas_compare}, within a diameter of $\sim10\mathrm{kpc}$ (see text for details). Measured velocities have been corrected for inclination to show the intrinsic rotational velocity.}\label{figure:rotcurves_stars}
\end{figure*}

Finally, the line-of-sight velocities of all fitted spaxels are plotted against their velocity dispersions in Figure~\ref{figure:kinematics_hist}. The x axis tries to separate the disk of the galaxy from the stripped high velocity tails, while the y axis can be used to distinguish between collapsing and heated/shocked gas. The plot shows a bimodal distribution of line-of-sight velocities within the disk of the galaxy visible in the lower left region of the plot ($\Delta v \lesssim 200 km s^{-1}$ and $\sigma v \lesssim 70 km s^{-1}$), with a visible skew to positive velocities due to the influence of strong line-of-sight gas stripping. At velocities of $\lesssim 350$ km~s$^{-1}$, the stripped gas follows a steep increase in velocity dispersion with increasing velocity, owing to the onset of turbulent motion within the stripped gas as it moves away from the galaxy. Above $350$km~s$^{-1}$ the trend reverses as the dispersion falls, likely  due to the cooling of stripped gas recollapsing.

Although the unique direction of stripping and viewing angle of JO201 makes the comparison with other jellyfish galaxies difficult, our results are consistent with observations of nearby jellyfish that show a structured distribution of the stripped gas composed of compact blobs and diffuse filaments \citep[e.g.][]{Boselli2016}, as well as simulated RPS events \citep{Tonnesen2010}. Moreover, previous    studies of the dynamics of jellyfish galaxies have shown that the stripped gas (primarily in the plane of the sky)  preserves the rotation of the galaxy for a while  \citep[e.g. up to $\sim 20\mathrm{kpc}$ in ESO137-001;][]{Fumagalli2014}, and then turbulent motion takes over.
In JO201 we observe a similar phenomenon, but primarily along the line of sight. 


\subsection{Rotation Curves} \label{sec:RC}

To directly compare the distribution of the  stellar component and the $\mathrm{H}\alpha$ gas in velocity space across the galaxy, we construct rotation curves. 

The red curve in  Figure~\ref{figure:rotcurves_stars} shows the  rotation curve of the stellar component,  constructed by taking the median velocities in a strip approximately 10 arcseconds wide along the disk's kinematic major axis ($2.5 \deg$ from north to east). The $\mathrm{H}\alpha$ emission-line velocity map is plotted on top, in strips parallel to the direction of the stellar kinematic major axis, to show the overall velocity structure of the gas (black lines), as well as the spread within individual regions undergoing stripping. 
Velocities were corrected for galaxy inclination ($i=54\deg$ from face-on) by dividing the observed velocities by $\sin i$.

The offset between the stellar and gaseous components of the disk, first seen in Section~\ref{sec:kinematics}, is clearly visible in Figure~\ref{figure:rotcurves_stars}. 
A large fraction of the ionised gas is  redshifted with respect to the stellar disk as a consequence of strong ram-pressure drag. 
Interestingly, many of the individually stripped clumps of material (groups of black lines separated from the main body of the galaxy)  seem to preserve the overall gradient of the rotation curve of the galaxy, although redshifted.

To test whether the high-velocity stripped gas is bound to the galaxy, we computed an upper limit to the escape velocity, $v_{esc}$, using the potential of a thin exponential disk, with the form presented in equation 2.165 of \citet{BinneyandTremaine}: 
\begin{equation}
\Phi(r,0) = -\pi G \Sigma_{0} r \left[I_0(y)K_1(y) - I_1(y)K_0(y)\right]
\end{equation}
where $y=r/2r_d$, $I_n, K_n$ are modified Bessel functions, $r_d$  the disk scale length, and $\Sigma_{0}$ the central surface density of a thin exponential disk, calculated from the dynamical mass as $\Sigma_0=M_{dyn}/2\pi r_d^2$. 
To determine $M_{dyn}$ we followed equation 15 of \citet{vandenBosch2002}, using the maximum velocity values of the stellar rotation curve (${\sim} 190$km~s$^{-1}$ after correcting for an inclination of $54^\circ$, calculated from an axis ratio of ${\sim} 0.6$) and an I-band disk scale length 
(we used $5.56 \mathrm{kpc}$ as in Section~\ref{sec:env}, as this value is consistent with fits made to the stellar continuum using the MUSE datacube). 
The resulting dynamical mass was found to be $M_{dyn}= 5.10\times 10^{11} M_{\odot}$.
The calculated escape velocity ($v^2_{esc}(r) = -2 \Phi(r)$) is an upper limit, as it does not take into account the vertical extent of the disk. Moreover, we note that $v_{esc}(r)$ should ideally be compared with the magnitude of the full velocity vector of the gas, while our observations are limited to the line-of-sight velocity component. 
We plot $v_{esc}(r)$ as a dashed blue curve on top of the stellar rotation curve  of the galaxy in
Figure~\ref{figure:rotcurves_stars}. Because it only makes sense to compare absolute values of $\Delta v$ with $v_{esc}$, for $r_{maj}>0$ (where the stellar rotation curve has $\Delta v <0$) we have plotted $v_{esc}+2\Delta v$ instead\footnote{This accounts for the difference in using $\Delta v$ rather than $|\Delta v|$. The plotted function, p(r) should obey the relation $v_{esc}-|\Delta v| = p(r) - \Delta v$.}. The plotted line helps distinguished several clumps of stripped gas that are unbound from the disk. These are located in the outer parts of the disk lying above the line.

\begin{figure}\centering
\includegraphics[width=0.5\textwidth]{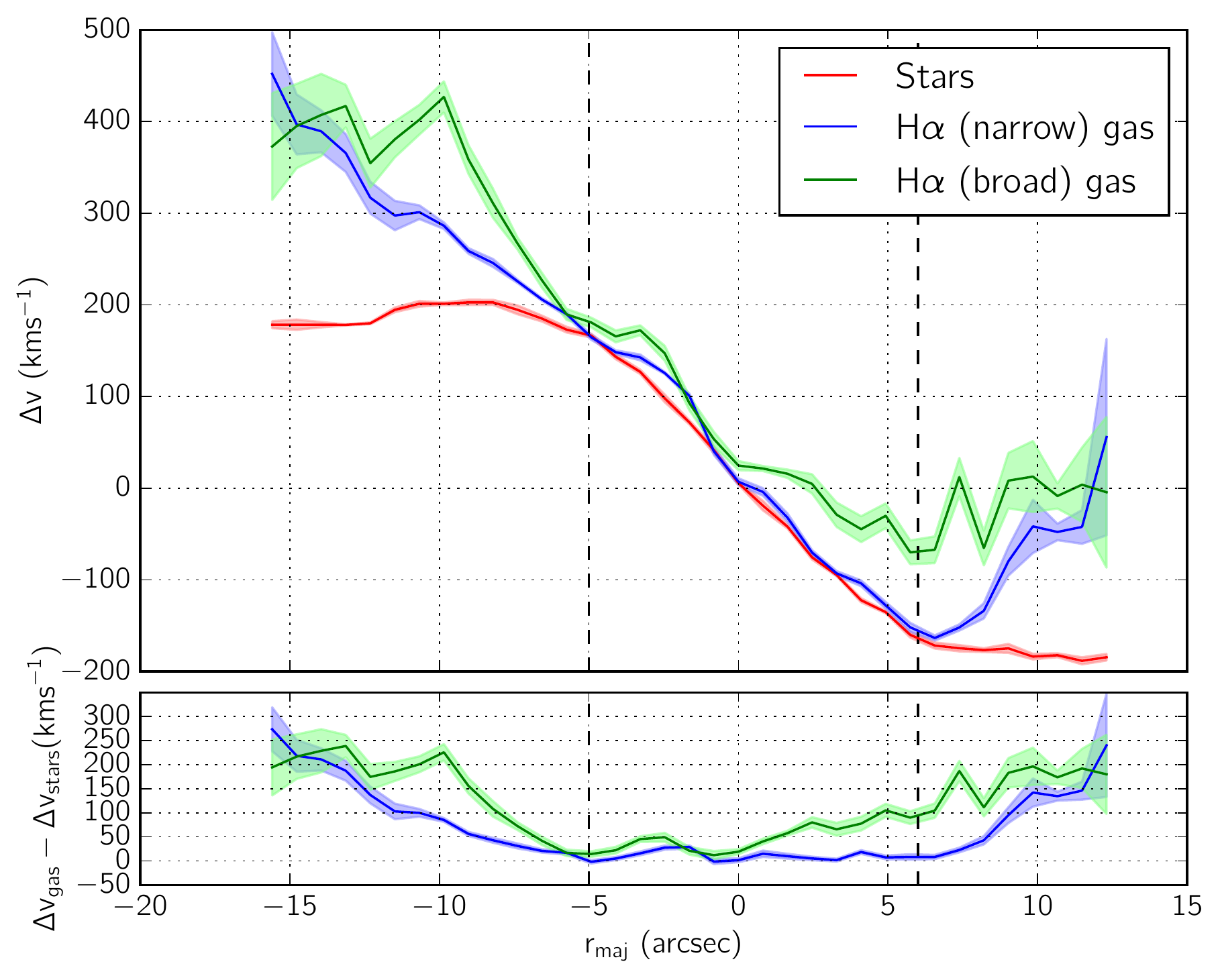}
\caption{Rotation curves of JO201, measured within a $\sim$10 arcsecond-wide strip along the kinematic major axis $\mathrm{r}_\mathrm{maj}$ of the stellar disk. Inclination-corrected median stellar velocities (obtained from the absorption line kinematics) are shown in red, along with the median velocity of the ionized gas, separated into narrow (blue), and broad component (green) of the H$\alpha$ emission-line fits. Standard deviations of bootstrap medians are shown for both. 
Dashed lines enclose the region of the disk less affected by ram-pressure stripping ($r_{maj}\lesssim 6 \mathrm{kpc}$).}
\label{figure:starsgas_compare}
\end{figure}

Figure~\ref{figure:starsgas_compare} presents inclination-corrected rotation curves of the stars and the ionised gas, separated into narrow and broad components of the $\mathrm{H}\alpha$ emission.
All of the curves were constructed by taking the median velocities in a strip approximately 10 arcseconds wide along the disk's kinematic major axis.
In the lower part of the figure, the residuals of the narrow and broad gaseous components from the stellar component are shown.

The rotation curves highlight the existing offset between the stellar and gaseous components of the disk.
In particular, the rotation curve of the gaseous component shows an upward turn at the edges of the stellar disk, $|r_{maj}|\gtrsim $ 6~arcsec. The residuals show clearly the increase in the velocity offset with distance from the centre of the disk, with the broad component showing a more pronounced effect. Note that the difference between the blue and green lines in Figure~\ref{figure:starsgas_compare} reflects the velocity offset displayed in Figure~\ref{figure:veldiff}. 
The observed behaviour can be interpreted as a decrease in the gravitational binding of the gas at higher distances from the centre of the galaxy (see Section~\ref{sec:env}), that allows the gas to be more easily stripped from the edges of the disk.

In summary, Figures~\ref{figure:rotcurves_stars} and~\ref{figure:starsgas_compare}  highlight the intensity of the stripping experienced by JO201, with a redshifted ionised gas component with respect to the stellar disk. This is consistent with ram-pressure dragging the gas away from the fast-moving galaxy (and the observer). The incremental shift in velocity observed at larger distances from the centre of the galaxy indicates outside-in gas removal, and the shape of the asymmetric rotation curve of the gas is consistent with simulations of face-on RPS \citep{Kronberger2008}.

\section{Summary and Conclusions}\label{section:conclusion}

This paper is the second of a series presenting the first results from the ongoing GASP (GAs Stripping phenomena in galaxies with MUSE) survey at the VLT (Poggianti et al. submitted, Paper I), that is obtaining deep integral-field spectrocopy for over 100 galaxies with signatures of gas stripping (so-called ``jellyfish" galaxies). 
We focus on the distribution and kinematics of the stellar and ionised gas components of JO201, one of the most striking jellyfish galaxy in the  GASP survey and its parent sample \citep{2016AJ....151...78P}.

We start by studying the global environment of 
JO201 and find that it is part of a massive unrelaxed galaxy cluster (A85). Its large differential line-of-sight velocity with respect to the cluster, and (projected) proximity to the BCG, indicate JO201 is likely falling into the cluster for the first time, following a radial orbit mostly in the direction towards the observer. %
Moreover, a dynamical analysis utilizing all the cluster members suggests that JO201 is possibly falling into A85 within a small group of galaxies.

JO201 has been classified as a jellyfish galaxy candidate by \citet{2016AJ....151...78P} owing to the presence of optical tails in only one side of the galaxy, which  suggests that unilateral gas stripping is taking place. We  modelled the ram-pressure expected in A85 and find that, at the projected position and velocity of JO201, a galaxy of its mass and size must have lost $\sim 50\%$ of its total gas mass during its first passage though the cluster.
The amount of lost gas coincides remarkably well with the estimated gass loss computed by comparing the size of the stellar disk with the extent of the remaining $\mathrm{H}\alpha$ (ionised gas) disk. Our environmental analysis thus strongly support RPS as the principal mechanism removing gas from JO201 during its first passage though the dense ICM.

We analysed the MUSE datacubes (covering JO201's main body and outskirts) to study the distribution and kinematics of the stellar and ionised gas components of the galaxy, and find the following.  

\begin{enumerate}
\item There is significant $\mathrm{H}\alpha$ emission concentrated along the windward side of the galaxy (colliding with the ICM), as well as  tails in the opposite direction, composed of bright  $\mathrm{H}\alpha$ knots and diffuse emission. The tails extend ${\sim} 50$kpc away from the galaxy, and point towards the cluster centre.  
The observed (structured) distribution of the gas is consistent with RPS simulations of galaxies \citep{Tonnesen2010}. 
We identify a pattern in the distribution of stripped gas in the tails of JO201 that resembles the shape of wide spiral arms. If true, a possible explanation is that strong RPS can cause an unwinding of the outer spiral arms. Owing to the difficulty in reproducing realistic spiral arms in simulated galaxies, this effect has not (yet) been reported by simulations of RPS.   Observations of other jellyfish galaxies have not reported the effect either, but this is likely a consequence of the rare angle of observation of the JO201. We hope to explore and expand upon this idea in future GASP studies.

\item In some parts of the galaxy the $\mathrm{H}\alpha$ emission often displays a non-gaussian profile. 
We characterise the different types of line profiles within the galaxy and find that at the centre of the galaxy there is emission-line broadening caused by an active nucleus.
In the outer parts of the disk and stripped tails, the emission lines can be described by either a single narrow gaussian, or a narrow component with an offset (redward) broader component. These double-gaussian profiles result from intense gas stripping along the line of sight, that drags material away from the galaxy (in the opposite direction to the observer).  

\item After selecting the best fit (single vs. double-component gaussians) in each spaxel, we constructed a three-dimensional model of the H$\alpha$ emission in the galaxy that highlights the extent of the emission line wings in the redshift direction. 

\item From a kinematic analysis of the absorption features, we find that the velocity field of the stellar component is well reproduced by a smoothly rotating disk. This suggests that there are no gravitational perturbations such as tidal interactions acting on JO201. The velocity of the gaseous component follows that of the stars in the inner regions of the galaxy, but deviates significantly from the disk to the tails. In particular, much of the stripped material with H$\alpha$ emission resides at higher line-of-sight velocity with respect to the stellar disk. The trend, clearly seen in the rotation curves of the stellar and gaseous components, increases with distance from the galaxy centre, in agreement with simulations of face-on RPS \citep{Kronberger2008}. 

\item The velocity dispersion in the stripped diffuse gas  is high ($>100$~km~s$^{-1}$), while the star-forming blobs are kinematically cold (velocity dispersion $<40$~km~s$^{-1}$), 
suggestive of shock heating and gas compression respectively. 
\end{enumerate}

Overall, our results favour RPS as the main mechanism removing gas from JO201. This galaxy's trajectory in the cluster, together with our viewing angle,  offer a unique insight into the ram-pressure drag generated by the ICM, with most of the stripping happening along our line of sight as the galaxy plunges through the cluster towards the observer.  
Moreover, the MUSE observations reveal clear evidence for outside-inward RPS of the disk, with gas compression and shock heating occurring in the stripped tails. 
In a forthcoming paper (Bellhouse et~al.~in preparation) we will further investigate the sources of ionisation in this galaxy, along with an analysis of the gas mass loss, gas-phase metallicity, and ages of the stellar populations across this galaxy.

JO201 is a an extreme case of stripping of a massive galaxy ($\sim 10^{11} M_{\odot}$) in a massive ($\sim 10^{15} M_{\odot}$) cluster. Our results are consistent with previous studies of jellyfish galaxies in a variety of environments, and thus highlight the importance and wide range of action that RPS has on satellite galaxies.

\section*{}
We would like to thank the anonymous referee for helping improve the paper significantly.
This work is based on observations collected at the European Organisation for Astronomical Research in the Southern Hemisphere under ESO programme 196.B-0578.
This work made use of the \textsc{kubeviz} software which is publicly available at: \url{http://www.mpe.mpg.de/~dwilman/kubeviz/}
We thank Matteo Fossati and Dave Wilman for their invaluable help with \textsc{kubeviz}.
C.B. thanks Frederic Vogt for the advice on data visualisation and on producing and presenting 3D visualisations of emission lines.
We thank Graeme Candlish for useful discussions.
This work was co-funded under the Marie Curie Actions of the European Commission (FP7-COFUND).
We acknowledge financial support from PRIN-INAF 2014
J.F. acknowledges the financial support from UNAM-DGAPA-PAPIIT IA10+15 grant, M\'exico.
B.V. acknowledges the support from an Australian Research Council Discovery Early Career Researcher Award (PD0028506)
%

\bibliographystyle{apj}
\bibliography{references.bib}
\end{document}